%% file: EXO-14-014_temp.tex
\pdfoutput=1

\documentclass[11pt,twoside,a4paper,cmspaper,final,collab]{cms-tdr}
\def\svnVersion{341051}\def\cmsCernNoTag{CERN-EP/2016-032}\def\cmsCernDate{\today}\def\cmsMessage{Published in the Journal of High Energy Physics as \href{http://dx.doi.org/10.1007/JHEP04(2016)169}{\doi{10.1007/JHEP04(2016)169.}}}

\begin{document}\cmsNoteHeader{EXO-14-014}

\hyphenation{had-ron-i-za-tion}
\hyphenation{cal-or-i-me-ter}
\hyphenation{de-vices}
\RCS$HeadURL: svn+ssh://svn.cern.ch/reps/tdr2/papers/EXO-14-014/trunk/EXO-14-014.tex $
\RCS$Id: EXO-14-014.tex 331204 2016-03-08 08:09:00Z ghm $
\newlength\cmsFigWidth
\ifthenelse{\boolean{cms@external}}{\setlength\cmsFigWidth{0.85\columnwidth}}{\setlength\cmsFigWidth{0.4\textwidth}}
\ifthenelse{\boolean{cms@external}}{\providecommand{\cmsLeft}{top\xspace}}{\providecommand{\cmsLeft}{left\xspace}}
\ifthenelse{\boolean{cms@external}}{\providecommand{\cmsRight}{bottom\xspace}}{\providecommand{\cmsRight}{right\xspace}}

\newcommand{\N}{\ensuremath{\mathrm{N}}\xspace}
\newcommand{\VmN}{\ensuremath{V^{}_{\mu \N}}\xspace}
\newcommand{\VeN}{\ensuremath{V^{}_{\Pe \N}}\xspace}
\newcommand{\VlN}{\ensuremath{V^{}_{\ell \N}}\xspace}
\newcommand{\VmNst}{\ensuremath{V^{*}_{\mu \N}}\xspace}
\newcommand{\VeNst}{\ensuremath{V^{*}_{\Pe \N}}\xspace}
\newcommand{\VlNst}{\ensuremath{V^{*}_{\ell \N}}\xspace}
\newcommand{\mN}{\ensuremath{m_{\N}}\xspace}
\newcommand{\mW}{\ensuremath{m_{\PW}}\xspace}

\cmsNoteHeader{EXO-14-014}

\title{Search for heavy Majorana neutrinos in $\Pe^\pm \Pe^\pm$+ jets and $\Pe^\pm \mu^\pm$+ jets events in proton-proton collisions at $\sqrt{s} = 8\TeV$}

\date{\today}

\abstract{
A search is performed for heavy Majorana neutrinos (\N) decaying into a \PW boson and a lepton using the CMS detector at the 
Large Hadron Collider. A signature of two jets and either two same sign electrons or a same sign electron-muon 
pair is searched for using 19.7\fbinv of data collected during 2012 in 
proton-proton collisions at a centre-of-mass energy of 8\TeV. The data are found to be consistent with the expected 
standard model (SM) background and, in the context of a Type-1 seesaw mechanism, upper limits are set on the cross 
section times branching fraction for production of heavy Majorana neutrinos in the mass range between 40 and 500\GeV. 
The results are additionally interpreted as limits on the mixing between the heavy Majorana neutrinos and the SM 
neutrinos. In the mass range considered, the upper limits range between 0.00015-0.72 for $\abs{\VeN}^2$ and 
$6.6 \times 10^{-5}$-0.47 for $| \VeN \VmNst |^2 / (  | \VeN |^2 + | \VmN |^2 )$, where $\VlN$ is the mixing element describing 
the mixing of the heavy neutrino with the SM neutrino of flavour $\ell$. 
These limits are the most restrictive direct limits for heavy Majorana neutrino masses above 200\GeV. 
}

\hypersetup{%
pdfauthor={CMS Collaboration},
pdftitle={Search for heavy Majorana neutrinos in e+/- e+/- + jets and e+/- mu+/- + jets events in proton-proton collisions at sqrt(s) = 8 TeV},%
pdfsubject={CMS},%
pdfkeywords={CMS, physics}}

\maketitle

\section{Introduction}

The discovery of neutrino oscillations established that neutrinos have non-zero masses and hints at possible  
physics beyond the standard model (SM). Results from various neutrino oscillation experiments together with cosmological 
constraints imply very small neutrino masses~\cite{pdg}. The leading model that naturally generates light 
neutrino masses is the so-called ``seesaw" mechanism, which can be realized in several different 
schemes~\cite{seesawI_1, seesawI_2, seesawI_3, seesawI_4, seesawII_1, 
seesawII_2,seesawII_3,seesawII_4,seesawII_5,seesawII_6, seesawIII}. In the simplest model, the smallness 
of the observed neutrino masses is due to a heavy neutrino state N. In this model the SM neutrino mass is 
given by $m_\nu \sim y^2_\nu v^2/ \mN$, where $y_\nu$ is a Yukawa coupling, $v$ is the Higgs vacuum expectation value in 
the SM,  and $\mN$ is the mass of the heavy neutrino state. If the seesaw mechanism were to explain the 
masses of the known neutrinos, both the light and the heavy neutrinos would have to be Majorana particles, so processes 
that violate lepton number conservation by two units would be possible.  Therefore, searches for heavy Majorana neutrinos 
using hadron colliders are very important in resolving the nature of neutrinos and the origin of neutrino masses. 

In this paper a search for heavy Majorana neutrinos using a phenomenological 
approach~\cite{Maj_hadCol_1, Maj_hadCol_2, Maj_hadCol_3, Maj_hadCol_4, Maj_hadCol_5,NOTETao,NOTEATLAS,NOTETao2} is 
described.  A Type-1 seesaw model is considered based on Refs.~\cite{NOTETao,NOTEATLAS,NOTETao2}
with at least one heavy neutrino that mixes with the SM neutrinos, with $\mN$ and \VlN as free parameters 
of the model. Here \VlN is a mixing element describing the mixing between the heavy Majorana neutrino and the SM neutrino of flavour $\ell$. 

Previous direct searches for heavy Majorana neutrinos have been reported by the DELPHI~\cite{delphi} and L3~\cite{l3, l3_2001} 
collaborations at LEP. They searched for $\Pe^+ \Pe^- \rightarrow \N \nu_{\ell}$, where $\nu_{\ell}$ is 
any SM neutrino ($\ell = \Pe$, $\mu$, or $\tau$), from which they set limits on the mixing element squared $|V^{}_{\ell \N}|^2$. 
For $\ell = \mu , \tau$ the limits are set for $\mN < 90\GeV$, while for $\ell = \Pe$ the limits extend to $\mN < 200\GeV$.
Several experiments have obtained limits for low neutrino masses ($\mN < 5\GeV$), 
including the LHCb Collaboration~\cite{Aaij:2014aba} at the LHC, which set limits on the mixing of a heavy neutrino with 
a SM muon neutrino. The searches by L3, DELPHI, and LHCb include the possibility of a heavy-neutrino lifetime sufficiently long 
that the decay vertex is displaced from the interaction point, while in the search reported here it is assumed 
that the \N decays with no significant displacement of the vertex since in the mass range of this search ($\mN > 40\GeV$)
the theoretical decay length is less than $10^{-11}\unit{m}$~\cite{NOTETao2}.  

Precision electroweak measurements have been used to constrain the mixing elements 
\begin{linenomath}
\begin{equation}
\Omega_{\ell \ell'} 
=  \sum\limits_{j=1}^n V^{}_{\ell \N_j}V^{*}_{\ell' \N_j},
\end{equation}
\end{linenomath}
where 
$j$ runs over heavy neutrino flavour states~\cite{NOTEATLAS}, resulting in indirect 90\% confidence level limits of
\begin{linenomath}
\begin{equation}
\label{omega}
\Omega_{\Pe\Pe} = \sum_j | V^{}_{\Pe \N_j} |^2 < 0.003,~~~
\Omega_{\mu \mu} = \sum_j | V^{}_{\mu \N_j} |^2 < 0.003,~~~ 
\Omega_{\tau \tau} = \sum_j | V^{}_{\tau \N_j} |^2 < 0.006,
\end{equation} 
\end{linenomath}
which are independent of \mN~\cite{delAguila:2008pw}.
Further restrictions are set on the mixings from flavour changing neutral current processes. 
These bounds depend on the mass of the heavy neutrinos. For $\mN = 10\GeV$ the limit 
$|\Omega_{\Pe \mu}| =  | \sum_j  V^{}_{\Pe \N_j} V_{\mu \N_j}^* |  \leq 0.015$ was set, 
while for the case that $m^2_{\N_j} \gg \mW^2 \gg \abs{ V^{}_{\ell \N_j} }^2 m^2_{\N_j}$, a more stringent limit of
$|\Omega_{\Pe \mu}| \leq 0.0001$ was set~\cite{NOTETao2}. Additionally, for the mixing of the Majorana neutrino 
with the SM electron neutrino, the limits from neutrinoless double beta decay are~\cite{Aalseth:2004hb}
\begin{linenomath}
\begin{equation}
\label{eq:v0bb}
   \sum\limits_{j=1}^n | V^{}_{\Pe \N_j} |^2 \frac{1}{m_{N_{j}}} < 5 \times 10^{-8}\GeV^{-1} ,
\end{equation}
\end{linenomath}
where $j$ runs over heavy neutrino flavour states.
However, the neutrinoless double beta decay experiments can only set limits on mixing with first generation leptons.
Collider experiments on the other hand can also search for mixing with second and third generation fermions.
If $V^{}_{\Pe \N_j}$ saturates $\Omega_{\Pe\Pe}$ in Eq.~(\ref{omega}), the limit on \VeN from neutrinoless double beta decay can 
be satisfied either by demanding that \mN is beyond the TeV scale, or that there are cancellations among the different 
terms in Eq.~(\ref{eq:v0bb}), as may happen in certain models~\cite{ingleman}. Other models with heavy neutrinos 
have also been examined. The ATLAS and CMS collaborations at the LHC have reported limits on heavy Majorana neutrino 
production in the context of the Left-Right Symmetric Model~\cite{ATLAS:2012ak, Khachatryan:2014dka}. The ATLAS experiment also set 
limits based on an effective Lagrangian approach~\cite{ATLAS:2012ak}. 

Because of the Majorana nature of the heavy neutrino considered here, both opposite- and same-sign lepton pairs can 
be produced. This search concentrates on the same-sign dilepton signatures since these final states have very low 
SM backgrounds. In addition to these leptons, the Majorana neutrino also produces an accompanying \PW boson when it decays.
We search for \PW decays to two jets, as this allows reconstruction of the mass of the heavy 
neutrino without missing any transverse momentum associated with SM neutrinos.

\begin{figure}[htbp]
\begin{center}
\includegraphics[width=0.4\textwidth]{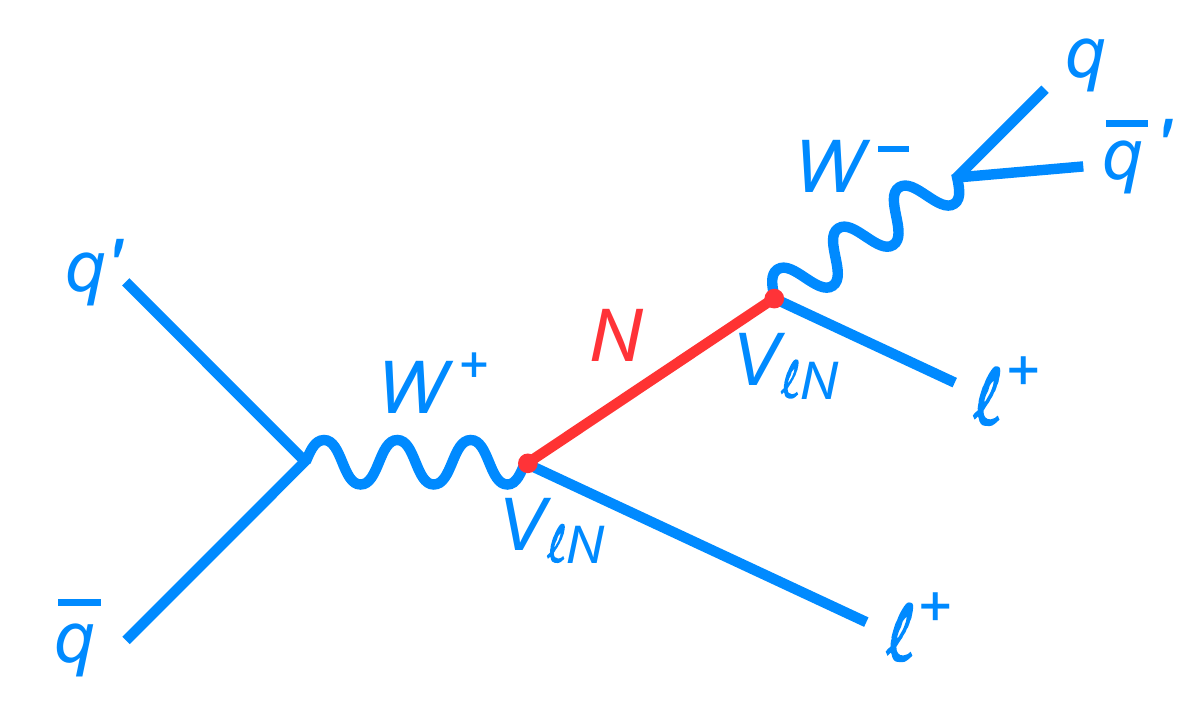}
\caption{The Feynman diagram for resonant production of a Majorana neutrino (\N).
         The charge-conjugate diagram results in a $\ell^- \ell^-\PAQq\PQq^\prime $ final state.}
\label{fig:feynman}
\end{center}
\end{figure}

The dominant production mode of the heavy neutrino under consideration is shown in Fig.~\ref{fig:feynman}. 
In this process the heavy Majorana neutrino is produced by $s$-channel production of a \PW boson, which decays via $\PW^+ 
\rightarrow \N\ell^+$. The \N can decay via $\N \rightarrow \PW^- \ell^+$ with 
$\PW^{-} \rightarrow\PQq\PAQq^\prime$, resulting in a $\ell^{+} \ell^{+} \PQq\PAQq^\prime$ 
final state. The charge-conjugate decay chain also contributes and results in a $\ell^-\ell^- \PAQq \PQq^\prime$ 
final state. In the this analysis, only $\ell = \Pe$~or~$\mu$ is considered. In a previous publication~\cite{CMS_NR_mu_2012} by the 
CMS Collaboration a search for heavy neutrinos in events with a dimuon final state was reported. In the present paper the 
search is expanded to include events with $\Pe^\pm \Pe^\pm \PQq\PAQq^\prime$ and 
$\Pe^\pm \mu^\pm \PQq\PAQq^\prime$ final states. These decay modes are referred to as the 
dielectron and electron-muon channels, respectively.  
The lowest order parton subprocess cross section $\hat \sigma (\hat s)$ for $\PQq \PAQq^\prime \to (\PW^\pm)^* \to \N \ell^\pm$ at 
a parton center-of-mass energy $\sqrt{\hat s}$ is given by 
is given~\cite{Pilaftsis} by:
\begin{linenomath}
\begin{equation}
        \hat \sigma( \hat s) = 
		\frac{\pi \alpha_W^2}{72 {\hat s}^2 \left[ \hat s - ( \mW - \frac{i}{2} \Gamma_{\PW})^2 \right] } 
		\abs{\VlN}^2 (\hat s - \mN^2)^2 (2 \hat s + \mN^2), 
\end{equation}
\end{linenomath}
where $\alpha_W$ is the weak coupling constant, and $\mW$ and $\Gamma_{\PW}$ are the \PW boson mass and width, respectively.

Observation of a $\ell^- \ell^{(\prime)-} \PQq\PAQq^\prime$ signature would constitute direct evidence 
of lepton number violation. The study of this process in different dilepton channels 
brings greater likelihood for the discovery of a Majorana neutrino, and constrains the mixing elements. The dielectron and electron-muon
channels allow constraints to be set on $|\VeN|^2$ and $|\VeN \VmNst |^2 / ( |\VeN|^2 + |\VmN|^2 )$, respectively.  

In a previous publication, a search for a heavy neutrino with mass less than 200\GeV was performed by the 
CMS Collaboration using dielectron and dimuon events at $\sqrt{s} = 7\TeV$~\cite{CMS_NR_2011}. The CMS experiment also 
searched for a heavy neutrino with mass up to 500\GeV using the dimuon channel at $\sqrt{s} = 8\TeV$~\cite{CMS_NR_mu_2012}.
The search has also been performed  by ATLAS using 8\TeV data for masses up to 500\GeV~\cite{ATLAS_NR_2012}. 
While these results constrain $|\VmN|^2$ and $|\VeN|^2$, there are no current direct limits on $|\VeN \VmNst |^2 / ( |\VeN|^2 + |\VmN|^2 )$. 

In this paper, an updated search for a heavy Majorana neutrino in the dielectron channel and a new search in the 
electron-muon channel are presented using CMS data collected in 2012 at $\sqrt{s} =  8\TeV$.  We search for events with either two 
electrons or  one electron and one muon, both of the same-sign electric charge, and in both cases at least two accompanying 
jets are required. Heavy neutrinos with a mass in the range 
of 40 to 500\GeV are considered. 

There are three potential sources of same-sign dilepton backgrounds: SM sources in which two prompt same-sign leptons are 
produced (e.g. $\PW\PZ$ production), events resulting from misidentified leptons, and opposite-sign dilepton events (e.g. from 
$\PZ\rightarrow \ell^{+} \ell^{-}$, $\PW\PW\rightarrow \ell^{+} \nu \ell^{-}\nu$) in which the 
charge of one of the leptons is mismeasured. The latter source is negligible for the electron-muon channel but 
is an important background in the dielectron channel.

\section{CMS detector and simulation}
\label{detector_sim}

The central feature of the CMS detector is a superconducting solenoid with an internal radius of 3\unit{m}. The solenoid 
provides a magnetic field of 3.8\unit{T}. Within the solenoid volume are a silicon pixel and strip tracker, a 
lead tungstate crystal electromagnetic calorimeter (ECAL), and a brass and scintillator hadron calorimeter, each 
composed of a barrel and two endcap sections. Muons are measured in gas ionization detectors embedded in the steel 
flux-return yoke outside the solenoid. Extensive forward calorimetry complements the coverage in pseudorapidity provided by the barrel and endcap 
detectors, where pseudorapidity is defined as $\eta = -\ln[\tan(\theta/2)]$. A two-tier triggering 
system selects the most interesting events for offline analysis. A more detailed description of the CMS detector, as 
well as definitions of the coordinate system used, can be found in Ref.~\cite{CMS_detector}. 

Several samples of simulated SM processes, 
which include a full treatment of the $\Pp\Pp$ collisions, were used. These samples include the full simulation of the CMS detector based 
on \GEANTfour~\cite{Geant4} and are reconstructed using the same CMS software as used for data. To ensure correct 
simulation of the number of additional interactions per bunch crossing (pileup), simulated events are mixed with 
multiple simulated  minimum bias events. Each simulated event is then weighted such that the distribution of pileup interactions in 
the simulation matches that in the data. 

A number of Monte Carlo (MC) event generators are used to simulate signal and background events: \ALPGEN~v2.14~\cite{ALPGEN}, 
\MADGRAPH~v5.1.3.30~\cite{Alwall:2011uj} and \PYTHIA~v6.4.22~\cite{pythia}. In 
order to simulate heavy Majorana neutrino events, a leading-order (LO) event generator described in 
Ref.~\cite{NOTEATLAS} is used, which was implemented in \ALPGEN~v2.14 with the CTEQ6M parton distribution functions 
(PDFs)~\cite{CTEQ6}. Parton showering and hadronization are simulated using \PYTHIA~v6.4.22.
The \PYTHIA~v6.4.22 generator is used to model the production of $\PW\PZ$ and $\PZ\PZ$ with fully leptonic final 
states. Events from double W-strahlung and double parton scatterings as well as triboson and $\PQt 
\PAQt$ plus boson ($\PQt \PAQt \PW$, $\PQt \PAQt \PZ$, 
$\PQt \PAQt \PW\PW$) are generated with \MADGRAPH v5.1.3.30. 

For production of same-sign same-flavour final states,  
$\Pp\Pp \rightarrow \N \ell^{\pm} \rightarrow \ell^{\pm} \ell^{\pm} \PQq\PAQq^\prime$, 
the heavy Majorana neutrino cross section is proportional to $\abs{\VlN}^2$ and is strongly dependent on \mN. 
The LO cross section at $\sqrt{s} = 8\TeV$ with $\abs{\VlN}^2 = 1$ has a value of 1515\unit{pb} for $\mN = 40\GeV$. 
The cross section is 3.56\unit{pb} for $\mN = 100\GeV$, and drops to 2.15\unit{fb} for $\mN = 500\GeV$~\cite{NOTEATLAS}. The LO 
cross section is scaled by a factor of 1.34 to account for higher-order corrections, based on the 
next-to-next-to-leading-order calculation in \textsc{FEWZ}~\cite{FEWZ1, FEWZ2} for $s$-channel $\PW^\prime$ 
production, which has the same production kinematics as the signal.

\section{Data sample and event selection}
\label{sec:selection}

The events used for analysis were selected from proton-proton collisions with an integrated luminosity of 
19.7\fbinv collected by the CMS detector in 2012.

\subsection{Event reconstruction}

This analysis uses reconstructed muons, electrons, jets, and a measurement of the missing transverse energy in the event.

Leptons, jets, and missing transverse energy in the event are reconstructed using a particle-flow 
event algorithm~\cite{pflow1, pflow2}. The missing transverse momentum vector is defined as the projection on the plane 
perpendicular to the beams of the negative vector sum of the momenta of all reconstructed particles in an event. Its 
magnitude is referred to as \MET. Jets are reconstructed by forming clusters of particle-flow candidates based on 
the anti-$\kt$ algorithm~\cite{antikt}, with a distance parameter of 0.5. Standard jet identification 
procedures are applied to suppress jets from calorimeter noise and beam halo.
Contributions from pileup are estimated using the
jet area method~\cite{pileup_1, pileup_2} and are subtracted from the jet \pt.
The energy of reconstructed jets is corrected based on 
the results of simulation and data studies~\cite{JetES}. 

Events used in this search are selected using dilepton triggers, requiring the highest-$\pt$ (``leading'') lepton to 
have $\pt > 17\GeV$ and the second highest-$\pt$ (``trailing'') lepton to have $\pt > 8\GeV$. The trigger efficiency, measured 
using events collected with hadronic triggers, is $0.96^{+0.04}_{-0.06}$ for $\Pe\Pe$ pairs where the trailing electron has $\pt > 30\GeV$, and 
decreases to $0.92 \pm 0.06$ for $15 < \pt <$~30\GeV. The $\Pe\mu$ trigger efficiency is $0.93 \pm 0.06$. 

Additional selections are performed after the trigger requirements to ensure the presence of identified leptons and 
jets. Events are first required to have a reconstructed $\Pp\Pp$ interaction vertex (primary vertex) identified as the 
reconstructed vertex with the largest value of $\Sigma \pt^{2}$ for its associated charged particle tracks 
reconstructed in the tracking detectors~\cite{vertex}. 

Electron (muon) candidates are required to have $|\eta|<2.5~(2.4)$ and must be consistent with originating from the 
primary vertex. Electron candidates must pass a number of identification requirements on the shower shape, track quality, 
and matching between the track and calorimeter energy deposit~\cite{cmsel}. Electrons must also not be consistent with originating from a photon 
conversion. The electrons must be well isolated from other activity in the event, which is ensured by 
requiring their relative isolation parameter ($I_{\text{rel}}$) to be less than 0.09 (0.05) in the barrel (endcap).
 Here $I_{\text{rel}}$ is defined as the scalar sum of transverse energy of the reconstructed particles present within 
$\Delta R = \sqrt{\smash[b]{(\Delta \eta)^{2} + (\Delta \phi)^{2} }} < 0.3$ of the candidate's direction, where $\phi$ is the azimuthal angle, excluding the candidate itself, 
divided by the lepton candidate \pt.  To ensure reliable determination of the electron charge, two independent measurements of the charge are considered. One method uses the curvature of the associated track measured in the silicon tracker. The other method compares the azimuthal angle between the vector joining the nominal interaction point and the ECAL cluster position, and the vector joining the nominal interaction point and the innermost hit of the track. The two methods are required to give consistent results.

Muon candidates are required to satisfy specific track quality and calorimeter deposition requirements~\cite{cmsmuon}. 
Muon candidates must also be isolated from other activity in the event, which is ensured by requiring $I_{\text{rel}}$ to be 
less than 0.05. Both electrons and muons are required to be well separated from jets, such that 
$\Delta R (\ell, \mathrm{jet}) > 0.4$. The lepton selection criteria are the same as those used in Ref.~\cite{SUSY_2012} 
except for the more stringent requirement on $I_{\text{rel}}$ for both electrons and muons. 

\subsection{Preselection criteria}

At the preselection stage, events are required to contain two same-sign leptons. The leading (trailing) lepton is required to have $\pt > 
20~(15)\GeV$. The invariant mass of the dilepton pair is required to be above 
$10\GeV$. For dielectron events a 20\GeV mass range centred on the \PZ boson mass is excluded
to reject background from \PZ boson decays in which one electron charge is mismeasured. In order to 
suppress backgrounds from diboson production, such as $\PW\PZ$, events with a third lepton identified using a looser set of 
requirements and with $\pt > 10\GeV$ are removed. 
Preselection events are required to have two or more jets that have $\pt > 20\GeV$ and
$|\eta| < 2.5$. To reduce backgrounds from top quark decays, events in which one of the jets is identified as originating from a bottom quark (b tagged) 
are rejected, where the medium working point of the combined secondary vertex 
tagger~\cite{btag} has been used. The b tagging efficiency is approximately 70\% with a misidentification probability for 
light-parton jets of 1.5\%.

\subsection{Selection criteria for signal region}

Depending on the $\mN$ hypothesis, signal events from heavy neutrino decays have different kinematic 
properties. In the low  mass search region ($\mN < \mW$), the \PW boson propagator that produces the heavy 
neutrino in Fig.~\ref{fig:feynman} is on-shell and the final state system of dileptons and two jets should have an 
invariant mass close to the \PW mass. In the high mass search region ($\mN > \mW$), the \PW boson propagator is 
off-shell but the \PW boson from the heavy neutrino decay is on-shell, so the invariant mass of the two jets from the \PW will be close to the \PW 
mass. Therefore, two different selection criteria were developed, depending on the heavy neutrino mass hypothesis to 
obtain the best sensitivity. For this analyses the simulated mass points are divided into low-mass (${<}90\GeV$) 
and high-mass (${\ge} 90\GeV$) search regions. 

The different selection requirements used for the low- and high-mass search regions are shown in Table~\ref{table:low_highmass}. In 
the low-mass search region, the following selections are imposed: $\MET < 30\GeV$; 
the invariant mass of the two leptons and two jets is required to be less than 200\GeV,
where the two jets chosen are those that give the invariant mass of the two leptons and two jets closest to $\mW$; 
the dilepton invariant mass is required to be greater than 10\GeV; 
the invariant mass of the two jets is required to be less than 120\GeV; 
and the leading jet must have $\pt > 20\GeV$.
In the high-mass search region the following selection cuts are used: $\MET < 35\GeV$; 
the invariant mass of the two leptons and two jets is required to be greater than 80\GeV,
the dilepton invariant mass is required to be greater than 15\GeV; 
the invariant mass of the two jets must satisfy $50 < m(\mathrm{jj}) < 110\GeV$;
and the leading jet must have $\pt > 30\GeV$.
In both the low- and high-mass search regions, the upper threshold on \MET suppresses SM background processes in which a \PW 
boson decays leptonically ($\PW \rightarrow \ell\nu$), including W+jet and $\PQt\PAQt$ 
production. 
\begin{table*}[tbh]
\topcaption{Selection requirements for the low- and high-mass signal regions.}
\label{table:low_highmass}
\begin{center}
\begin{tabular}{cccccc}
\hline 
\multirow{2}{*}{Region} & \MET & $m(\ell^{\pm} \ell^{\pm} \mathrm{jj})$ &  $m(\ell^{\pm} \ell^{\pm})$ & $m(\mathrm{jj})$ & $\pt^{\mathrm{j}_1}$ \\
&(\GeVns) & (\GeVns) &  (\GeVns) & (\GeVns)  & (\GeVns)\\
\hline
Low-Mass  &    ${<}30$  & ${<} 200$ & ${>} 10$  & ${<} 120$  & ${>} 20$ \\
High-Mass &    ${<}35$  &  ${>} 80$ & ${>} 15$  & 50-110 & ${>} 30$ \\
\hline
\end{tabular}
\end{center}
\end{table*}

After applying the above selection criteria, the signal significance is optimized for each mass hypothesis with six variables using a 
figure of merit~\cite{punzi} defined as $\epsilon_S / (1 + \delta B)$, where $\epsilon_S$ is the signal selection efficiency and 
$\delta B$ is the uncertainty in the estimated background. 
The six variables used to optimize the signal selection are: 
the transverse momentum of the leading lepton $\pt^{\ell_1}$; 
the transverse momentum of the trailing lepton $\pt^{\ell_2}$;
the transverse momentum of the leading jet $\pt^{\mathrm{j}_1}$;
the invariant mass of the two leptons and two selected jets $m(\ell \ell \mathrm{j j})$;
the invariant mass of the sub-leading lepton and two selected jets $m(\ell_{2} \mathrm{j j})$; 
and the invariant mass of the two leptons $m(\ell \ell)$. 
Table~\ref{table:optimize} shows the optimized 
selection requirements for the dielectron and electron-muon channels,  together with the overall signal acceptance. 
The overall signal acceptance ranges from 0.19-0.39\% for $\mN = 40\GeV$ to 
14-17\% for $\mN = 500\GeV$. 
Here, the lower acceptance at low $\mN$ is due to the selection requirements on the \pt
of the electrons and jets in a signal with very soft jets and electrons. 
The overall signal acceptance includes trigger efficiency, geometrical acceptance, and efficiencies of all selection criteria. 
\begin{table*}[h]
\topcaption{Selection requirements on discriminating variables determined by the optimization
for each Majorana neutrino mass point. The last column shows the overall signal acceptance. Different selection criteria
are used for low- and high-mass search regions. The ``---'' indicates that no selection requirement is made.}
\label{table:optimize}
  \begin{center}
  \footnotesize
  \begin{tabular}{cccccccc}
\hline 
$\mN$ & $p^{\ell_1}_{\mathrm{T}}$  & $\pt^{\ell_2}$ & $\pt^{\mathrm{j}_1}$ 
& $m(\ell^{\pm} \ell^{\pm} \mathrm{j j})$  & $m(\ell_{2} \mathrm{j j})$ & $m(\ell^{\pm} \ell^{\pm})$ & Acc. $\times$ Eff.\\
(\GeVns) & (\GeVns) &  (\GeVns)  & (\GeVns)& (\GeVns) & (\GeVns) &(\GeVns) & (\%) \\
\hline 
$\Pe\Pe$ channel: &&&&&&& \\ 
40        & ${>}20$    & ${>}15$  & ${>}20$  & 80-160  & ${<} 120$    &  10-60    & $0.19\pm0.01$  \\
50        & ${>}20$    & ${>}15$  & ${>}20$  & 80-160  & ${<} 120$    &  10-60    & $0.26\pm0.02$ \\
60        & ${>}20$    & ${>}15$  & ${>}20$  & 80-160  & ${<} 120$    &  10-60    & $0.22\pm0.01$ \\
70        & ${>}20$    & ${>}15$  & ${>}20$  & 80-160  & ${<} 120$    &  10-60    & $0.09\pm0.01$ \\
80        & ${>}20$    & ${>}15$  & ${>}20$  & 80-160  & ${<} 120$    &  10-60    & $0.32\pm0.02$ \\
90        & ${>}20$    & ${>}15$  & ${>}30$  & ${>}120$   & 60-120   & ${>}15$   & $0.46\pm0.03$ \\
100       & ${>}20$    & ${>}15$  & ${>}30$  & ${>}120$   & 80-120   & ${>}15$   &  $1.9\pm0.1$ \\
125       & ${>}25$    & ${>}25$  & ${>}30$  & ${>}140$   & 105-145  & ${>}15$   & $4.2\pm0.1$ \\
150       & ${>}40$    & ${>}25$  & ${>}30$  & ${>}195$   & 125-175  & ${>}15$   & $6.5\pm0.1$ \\
175       & ${>}45$    & ${>}30$  & ${>}30$  & ${>}235$   & 155-200  & ${>}15$   & $6.4\pm0.1$ \\
200       & ${>}65$    & ${>}40$  & ${>}30$  & ${>}280$   & 160-255  & ${>}15$   & $8.4\pm0.1$ \\
250       & ${>}110$   & ${>}40$  & ${>}40$  & ${>}300$   & ---          & ${>}15$   & $10.6\pm0.1$ \\
300       & ${>}120$   & ${>}40$  & ${>}40$  & ${>}320$   & ---           & ${>}15$   & $14.0\pm0.2$  \\
350       & ${>}120$   & ${>}40$  & ${>}40$  & ${>}360$   & ---          & ${>}15$   & $16.1\pm0.2$  \\
400       & ${>}120$   & ${>}40$  & ${>}40$  & ${>}360$   & ---          & ${>}15$   & $17.2\pm0.2$  \\
500       & ${>}120$   & ${>}40$  & ${>}40$  & ${>}360$   & ---          & ${>}15$   & $16.6\pm0.2$ \\
 $\Pe\mu$ channel: &&&&&&& \\ 
40           & ${>}20$    & ${>}15$    & ${>}20$  & 80-150    & ---    & ${>}10$    & $0.39\pm0.02$ \\
50           & ${>}20$    & ${>}15$    & ${>}20$  & 80-150    & ---    & ${>}10$    & $0.46\pm0.02$ \\
60           & ${>}20$    & ${>}15$    & ${>}20$  & 80-150    & ---    & ${>}10$    & $0.38\pm0.01$ \\
70           & ${>}20$    & ${>}15$    & ${>}20$  & 80-150    & ---    & ${>}10$    & $0.14\pm0.01$ \\
80           & ${>}25$    & ${>}15$    & ${>}20$  & 90-200    & ---    & ${>}10$    & $0.58\pm0.02$ \\
90           & ${>}40$    & ${>}15$   & ${>}30$   & ${>}120$     & ${<} 130$   & ${>}45$    &  $0.57\pm0.02$ \\
100          & ${>}40$    & ${>}30$   & ${>}30$   & ${>}130$     & ${<} 135$   & ${>}45$   & $1.71\pm0.04$ \\
125          & ${>}40$    & ${>}30$   & ${>}30$   & ${>}140$     & ${<} 160$   & ${>}45$   & $5.2\pm0.1$ \\
150          & ${>}45$    & ${>}30$   & ${>}30$   & ${>}150$     & ${<} 230$   & ${>}45$  & $9.5\pm0.1$ \\
175          & ${>}60$    & ${>}35$   & ${>}35$   & ${>}170$     & ${<} 240$   & ${>}45$   & $10.9\pm0.1$ \\
200          & ${>}75$    & ${>}35$   & ${>}35$   & ${>}200$     & ${<} 330$   & ${>}45$  & $11.9\pm0.1$ \\
250          & ${>}80$    & ${>}40$   & ${>}35$   & ${>}260$     & ${<} 390$   & ${>}45$   & $15.6\pm0.1$ \\
300          & ${>}110$   & ${>}40$   & ${>}35$   & ${>}310$     & ${<} 490$   & ${>}45$   & $16.0\pm0.1$  \\
350          & ${>}110$   & ${>}40$   & ${>}35$   & ${>}360$     & ${<} 550$   & ${>}45$  & $16.1\pm0.1$  \\
400          & ${>}120$   & ${>}40$   & ${>}35$   & ${>}380$     & ${<} 600$   & ${>}45$  & $16.2\pm0.1$  \\
500          & ${>}120$   & ${>}40$   & ${>}35$   & ${>}380$     & ${<} 700$   & ${>} 45$   & $14.1\pm0.1$ \\
\hline
\end{tabular}
\end{center}
\end{table*}

\section{Background estimation} 

\subsection{Background from prompt same-sign leptons}

Background events that result in two genuine, prompt leptons with the same charge are referred to as the prompt lepton 
background. This background is estimated using simulation. The largest 
contribution comes from $\PW\PZ$ and $\PZ\PZ$ events. Events from double W-strahlung and double parton scattering are also considered,
as well as triboson and $\PQt \PAQt$ plus boson ($\PQt \PAQt \PW$, $\PQt 
\PAQt \PZ$, $\PQt \PAQt \PW\PW$) production. Other rare processes include Higgs boson events in 
which the Higgs boson decays into neutral bosons ($\PH \to \PZ\PZ$), or 
$\PH \to \PW\PW$  contributions resulting from VH and $\PQt\PAQt 
\PH$ production. Backgrounds from $\PW\PZ$ 
and $\PZ\PZ$ events are estimated using the \PYTHIA~v6.4.22 generator, normalized to the next-to-leading order cross section, 
and \MADGRAPH~v5.1.3.30 for the remaining processes. 

\subsection{Background from misidentified leptons}

The most important background source originates from events containing objects misidentified as prompt leptons. These 
 originate from B hadron decays, light-quark or gluon jets as well as from photon conversions, and are 
typically not well isolated. The main components of this background are: multi-jet and $\gamma$+jet production, in which one or more jets are 
misidentified as leptons; $\PW (\rightarrow \ell\nu)$ + jets events, in which one of the jets is 
misidentified as a lepton; and $\PQt\PAQt$ decays, in which one of the top quark decays yields a prompt 
isolated lepton $(\PQt \rightarrow \PW\PQb \rightarrow \ell\nu \PQb)$, 
and the other lepton of same charge arises from a b quark decay or a jet misidentified as an isolated prompt lepton. 

Misidentified leptons are the dominant background for the low-mass search region. The simulation is not reliable in estimating
this background for several reasons, including limited statistical precision (due to the small 
probability of a jet to be misidentified as a lepton) and inexact modeling of the parton showering process. Therefore, 
these backgrounds are estimated using control samples from collision data as described below. 

An independent data sample enriched in multi-jet events (the ``measurement'' sample) is 
used to calculate the probability for a jet that passes minimal lepton selection requirements (``loose leptons'') to 
also pass the more stringent requirements used to define leptons selected in the full selection (``tight leptons''). 
This misidentification probability for the loose lepton is determined as a function of the lepton transverse momentum 
and pseudorapidity. 
This probability is used as a weight in the calculation of the background in events that pass all the signal 
selections except that one or both of the leptons fail the tight selection criteria, but pass the loose selection criteria. 
This sample is referred to as the ``application'' sample.

The misidentification probability is applied to the application sample by counting the number of events in which one 
lepton passes the tight selection, while the other lepton fails the tight selection but passes the loose selection 
($N_{n\bar{n}}$), and the number of events in which both leptons fail the tight selection, but pass the loose criteria 
($N_{\bar{n}\bar{n}}$). The total contribution to the signal sample (i.e. the number of events when both leptons pass 
the tight selection, $N_{nn}$), is then obtained by weighting events of type $n\bar{n}$ and $\bar{n}\bar{n}$ by the 
appropriate misidentification probability factors. To account for double counting a correction is made for $\bar{n}\bar{n}$ events 
that can also be $n\bar{n}$. 

The measurement sample is selected by requiring a loose lepton 
and a jet, resulting in events that are mostly dijet events with one jet containing a lepton. Since prompt leptons from 
\PW or \PZ decays tend to have high probability to pass the tight lepton requirements, any contamination of these prompt 
leptons in the measurement sample can bias the misidentification probabilities. To prevent this, a number of cuts are 
applied to remove these prompt leptons: only one lepton is allowed and upper thresholds on missing transverse energy 
($\MET < 20\GeV$) and the transverse mass ($m_{\mathrm{T}} < 25\GeV$) are applied, where $m_{\mathrm{T}}$ is calculated using the 
lepton \pt and \MET. These requirements suppress contamination from \PW and \PZ boson decays. The
loose lepton and jet are also required to be separated in azimuth by $\Delta \phi > 2.5$. 
This jet is used as a tag and the loose lepton as a probe used to determine the misidentification probability. 
The transverse energy of the tag jet is an essential ingredient to calibrate the characteristics of the probe (loose lepton). 
It is determined from MC simulations that the signal region is well modelled by a requirement of $\pt > 40\GeV$ on the tag jet, 
and therefore this is used to select the measurement sample. 

Loose electrons (muons) are defined by relaxing the standard identification requirements as follows:
the isolation requirement is relaxed from $I_{\text{rel}} < 0.09/0.05~(0.05/0.05)$ in the barrel (endcap) regions to $I_{\text{rel}} < 0.6~(0.4)$;
the transverse impact parameter of the lepton track is relaxed from ${<}0.10~(0.005)$\unit{mm} to ${<}10~(2)$\unit{mm};
for muons, the $\chi^2$ per degree of freedom of the muon track fit is relaxed from 10 to 50.

The method used to estimate the background from misidentified leptons is evaluated by checking the procedure using 
simulated event samples in which the true origin of the leptons, either from \PW or \PZ boson decays or from a quark decay, 
is known. The misidentification probabilities are obtained from multi-jet events (in simulation) and are used to 
estimate the misidentified lepton backgrounds in $\PW$ + jets events, as well as in an independent multi-jet simulated sample. 
This check was done using an inclusive selection of two leptons and two jets. 
For a sample of multi-jet events with with an integrated luminosity of 0.38\fbinv,
the number of true misidentified leptons is 4 in the dielectron channel, which agrees with the prediction of $5.0 \pm 2.1$. 
For the $\PW$ + jets simulation with an integrated luminosity of 1.54\fbinv, the true numbers of events with a misidentified lepton in the dielectron 
and electron-muon channels are found to be 18 and 26, respectively, while the predicted numbers are 
$22.1 \pm 7.2$ and $25.3\pm 8.8$. Thus, the predicted backgrounds agree with the expectations within the 
uncertainties. 

\subsection{Background from opposite-sign leptons}

To estimate backgrounds due to charge mismeasurement, the probability of mismeasuring the lepton charge is considered. The 
background due to mismeasurement of the muon charge was determined from simulation and from studies with cosmic ray muon data and found to be negligible
in the \pt range of interest in this search. Therefore, only mismeasurement of 
the electron charge is considered. 

The probability for mismeasuring the charge of a prompt electron is obtained from a simulation of $\PZ\rightarrow 
\Pe \Pe$ events and is parametrized as a function of $1/\pt^{\Pe}$ separately for electrons in the 
barrel and endcap calorimeters.  The average electron mismeasurement probability is found to be $(2.4\pm 0.3) 
\times 10^{-5}$ in the central ECAL barrel region ($|\eta| < 0.9$), $(1.1\pm 0.1) \times 10^{-4}$ at larger pseudorapidities in the ECAL barrel 
region ($0.9 < |\eta| < 1.5$), and $(3.2\pm 0.2) \times 10^{-4}$ in the ECAL endcap region ($1.5 < |\eta| < 2.5$). The 
charge mismeasurement probabilities are then validated, separately for barrel and endcap. To validate the charge 
mismeasurement probability for the barrel (endcap) region, a control sample of $\PZ \rightarrow 
\Pe \Pe$ events in the data is selected, requiring both electrons to pass through the barrel 
(endcap) region and requiring the invariant mass of the electron pair to be between 76 and 106\GeV. The difference 
between the observed and predicted numbers of $\Pe^{\pm}\Pe^{\pm}$ events is used as a scale factor to 
account for mismodeling in the simulation. The scale factors in the barrel and endcap are found to be $1.22 \pm 0.13$ 
and $1.40 \pm 0.21$ respectively. 

To validate the combined charge mismeasurement probability and scale factors, a control sample of $\PZ 
\rightarrow \Pe \Pe$ events in the data is again selected as described above but here 
requiring that one electron passes through the endcap and the other passes through the barrel region. The 
difference in the predicted and observed numbers of $\Pe^{\pm}\Pe^{\pm}$ events in this sample is 11\%. 
The same procedure is performed using $\PZ \rightarrow \Pe \Pe$ events in the data but 
requiring that the event has only one jet, obtaining agreement within 10\% between predicted background and data. 

To estimate the background in the dielectron and electron-muon channels, a weight of $W_{\mathrm{cm}}$ is applied to 
data events with all signal region cuts applied, except that here the leptons are required to be oppositely charged. Here, 
$W_{\mathrm{cm}}$ is given by $W_{\mathrm{cm}} = w_{\mathrm{cm_1}} /(1-w_{\mathrm{cm_1}}) + w_{\mathrm{cm_2}} /(1-w_{\mathrm{cm_2}})$, 
where $w_{\mathrm{cm_{1(2)}}}$ is the probability for the leading (trailing) electron charge to be mismeasured.

\subsection{Validation of background estimates}
\label{sec:bkg_valid}

To test the validity of the background estimation method, two signal free-control regions in data are defined. The background 
estimation method is applied in these regions and the result is compared with the observed yields. The control regions 
in the two mass ranges are defined as follows. In both the low-mass range ($40 < \mN < 90\GeV$) and the high-mass 
range ($\mN \geq 90\GeV$) the control region selection is the same as the signal selection without the final optimized 
selections but with either $\MET > 50\GeV$ or one or more jets that are b-tagged. 

The numbers of predicted and observed background events in the low- and high-mass control regions are shown in Table~\ref{tab:bkg_val}.
The misidentified lepton background accounts for about 2/3 of the total background in both regions. 
In both regions the predictions are in agreement with the observations within the systematic uncertainty
described in Section~\ref{sec:syst}, which is dominated by
the 35-40\% uncertainty in the misidentified lepton background.
The observed distributions of all relevant observables also agree with the predictions, within uncertainties.
\begin{table}[hptb]
\topcaption{Observed event yields and estimated backgrounds in the low- and high-mass control regions. 
            The uncertainty in the background yield is the sum in quadrature of the statistical and systematic uncertainties.}
\begin{center}
\begin{tabular}{c c c c}
\hline
Channel & Region &   Estimated Background & Observed \\
\hline
$\Pe\Pe$  & Low-mass  &  \phantom{0}21.4 $\pm$ 6.7\phantom{0}   & 18 \\
$\Pe\Pe$  & High-mass & \phantom{0}53.8 $\pm$ 15.3     & 44 \\
\hline
 $\Pe\mu$  & Low-mass  & \phantom{0}85.3 $\pm$ 21.8 & 68 \\
 $\Pe\mu$  & High-mass & 145.7 $\pm$ 35.2  & 119 \\
\hline
\end{tabular}
\label{tab:bkg_val}
\end{center}
\end{table}

\section{Systematic uncertainties}
\label{sec:syst}

The background estimate and signal efficiencies are subject to a number of systematic uncertainties.
The relative size of these uncertainties are listed in Table~\ref{tab:sys}.
A summary of the contribution of each systematic uncertainty relative to the signal or background estimate for two mass
points, $\mN = 100$ and $500\GeV$, is shown in Table~\ref{tab:sys_rel}.
\begin{table}[hptb]
\topcaption{Summary of the relative systematic uncertainties in heavy Majorana neutrino signal yields and the background from 
prompt same-sign leptons, both estimated from simulation. The relative systematic uncertainties 
assigned to the data-driven backgrounds for the misidentified lepton background and mismeasured charge background are also shown.
The uncertainties are given for the low-mass (high-mass) selections.}
\begin{center}
\begin{tabular}{lcccc}
\hline
\multirow{2}{*}{Channel / Source}                                &     $\Pe\Pe$ signal &  $\Pe\Pe$ bkgd.  &  $\Pe\mu$ signal &   $\Pe\mu$ bkgd. \\
                                                &    ($\%$)   & ($\%$)    & ($\%$) & ($\%$) \\
\hline
\underline{Simulation}: & & & & \\
SM cross section                                &  ---          & 9-25 (9-25)            & ---             & 9-25 (9-25)   \\
Jet energy scale                                & 6-8 (1-3)  & 5 (7)                      & 4-8 (1-2)   & 8 (7) \\
Jet energy resolution                         & 3-7 (2-3)  & 10 (7)\phantom{0} & 3-10 (2-3)\phantom{0} & 10 (6)\phantom{0} \\
Event pileup                                      & 2-3 (0-2)  & 4 (1)                       & 2-3 (0-2)   & 3 (2) \\
Unclustered energy                           & 1-3 (1-2)  & 4 (5)                       & 1-3 (1-2)   & 5 (1) \\
Integrated luminosity                         & 2.6 (2.6)  & 2.6 (2.6)                 & 2.6 (2.6)   & 2.6 (2.6) \\
Lepton selection                                & 2 (2)        & 2 (2)                       & 2 (2)         & 2 (2)\\
Trigger selection                                & 6 (6)        & 6 (6)                       & 6 (6)        & 6 (6) \\
b tagging                                            & 0-1 (1-2) & 2 (1)                       & 0-1 (1-2)  & 1 (1) \\
PDF (shape)                                      & 2.0 (2.0)  & ---                          & 2.0 (2.0) & --- \\
PDF (rate)                                          & 3.5 (3.5) & ---                           & 3.5 (3.5) & --- \\
Renormalization / factorization scales          & 8-10 (1-6)\phantom{0} & --- & 8-10 (1-6)\phantom{0} & --- \\
Signal MC statistical uncertainty               & 5-15 (1-6)\phantom{0} & --- & 3-7 (1-3) & ---\\
\underline{Data-Driven}: & & & & \\
Misidentified leptons 	&           ---  &  40 (40)  & --- & 35 (35) \\
Mismeasured charge  	&            --- &  12 (12)  &  --- &  12 (12) \\
\hline
\end{tabular}
\label{tab:sys}
\end{center}
\end{table}
\begin{table}[hptb]
\topcaption{Summary of contributions to the systematic uncertainty related to the prompt same-sign leptons background, misidentified lepton background, and  
mismeasured charge background on the total background uncertainty for the case of 
\mN= 100 and 500\GeV.}
\begin{center}
\begin{tabular}{ccccc}
\hline
\multirow{2}{*}{Channel} &   \mN  & Prompt bkgd. & Misid. bkgd. & Charge mismeas. bkgd.  \\
        &  (\GeVns)  & ($\%$) & ($\%$) & ($\%$) \\
\hline
$\Pe\Pe$   & 100 & \phantom{0}0.4 & 99.4  & 0.2   \\
$\Pe\Pe$   & 500 & \phantom{0}2.8 & 95.2  & 2.0   \\
\hline                                                        
 $\Pe\mu$   & 100  & \phantom{0}9.3  & 90.7  & 0.0 \\          
 $\Pe\mu$	 & 500  & 15.5 & 84.5  & 0.0 \\
\hline

\end{tabular}
\label{tab:sys_rel}
\end{center}
\end{table}

\subsection{Background uncertainties} 

The main sources of systematic uncertainties are associated with the background estimates. The largest uncertainty is 
that related to the misidentified lepton background. The overall systematic uncertainty in this 
background is determined by varying the background estimate with respect to the isolation requirement for the 
loose leptons and by varying the $\pt$ requirement for the tag jet. Increasing and decreasing the $\pt$ requirement 
for the tag jet changes the $\pt$ spectrum of the recoiling lepton in the event and is found to have the largest 
impact on the background level. As a result, the 35-40\% overall systematic uncertainty in the misidentified lepton 
background estimate is dominated by the $\pt$ requirement on the tag jet. 

For backgrounds from mismeasured electron charge an overall systematic uncertainty is assigned to the background of $12\%$. 
This is obtained by taking the weighted average of the uncertainties in the two scale factors in preselection events.
This uncertainty covers the difference between the predicted and observed numbers of events in both data closure tests.

\subsection{Simulation uncertainties}

The systematic uncertainties in the normalization of background from  prompt same-sign leptons are $12\%$ for $\PW\PZ$ 
and $9\%$ for $\PZ\PZ$~\cite{CMS_ZZ}. 
For the other processes the uncertainty is 25\%, determined by varying the renormalization and factorization scales 
from the nominal value of $Q^2$ to $4Q^2$ and $Q^2/4$, and following the PDF4LHC 
recommendations~\cite{Alekhin:2011sk,Botje:2011sn} to estimate the uncertainty due to the choice of PDFs. The overall 
systematic uncertainty in the prompt lepton background, including the sources discussed below, is 19-21\% for the low-mass selection 
and 18-19\% for the high-mass selection, depending on the channel. 
To evaluate the uncertainty due to imperfect knowledge of the integrated luminosity~\cite{CMS-PAS-LUM-13-001}, jet 
energy scale~\cite{JetES}, jet energy resolution~\cite{JetES}, b tagging~\cite{btag}, 
lepton trigger and selection efficiency, as well as the uncertainty in the cross section for minimum bias production 
used in the pileup reweighting procedure in simulation, the input value of each parameter is changed by $\pm1$ standard 
deviation from its central value. Energy not clustered in the detector affects the overall \MET 
scale resulting in an uncertainty in the event yield due to the requirement on \MET. Additional uncertainties in 
the heavy Majorana neutrino signal estimate arise from the choice of PDFs and renormalization and factorization scales 
used in the \ALPGEN~v2.14 MC event generator. These were also determined by varying the renormalization and factorization scales 
from the nominal value of $Q^2$ to $4Q^2$ and $Q^2/4$, and by following the PDF4LHC 
recommendations.

\section{Results}

\begin{figure}[tbph]
\begin{center}
\includegraphics[width=0.45\textwidth]{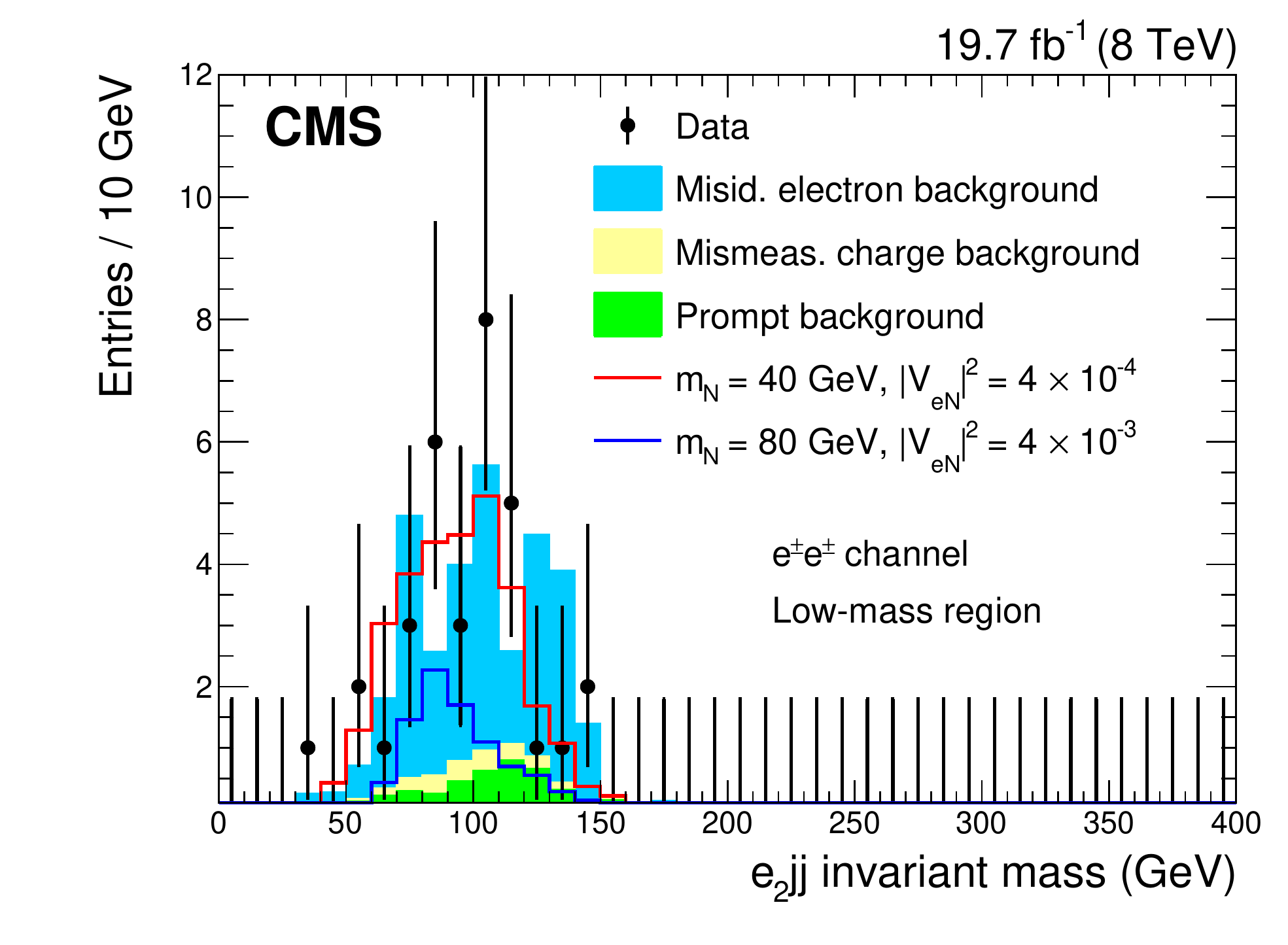} \includegraphics[width=0.45\textwidth]{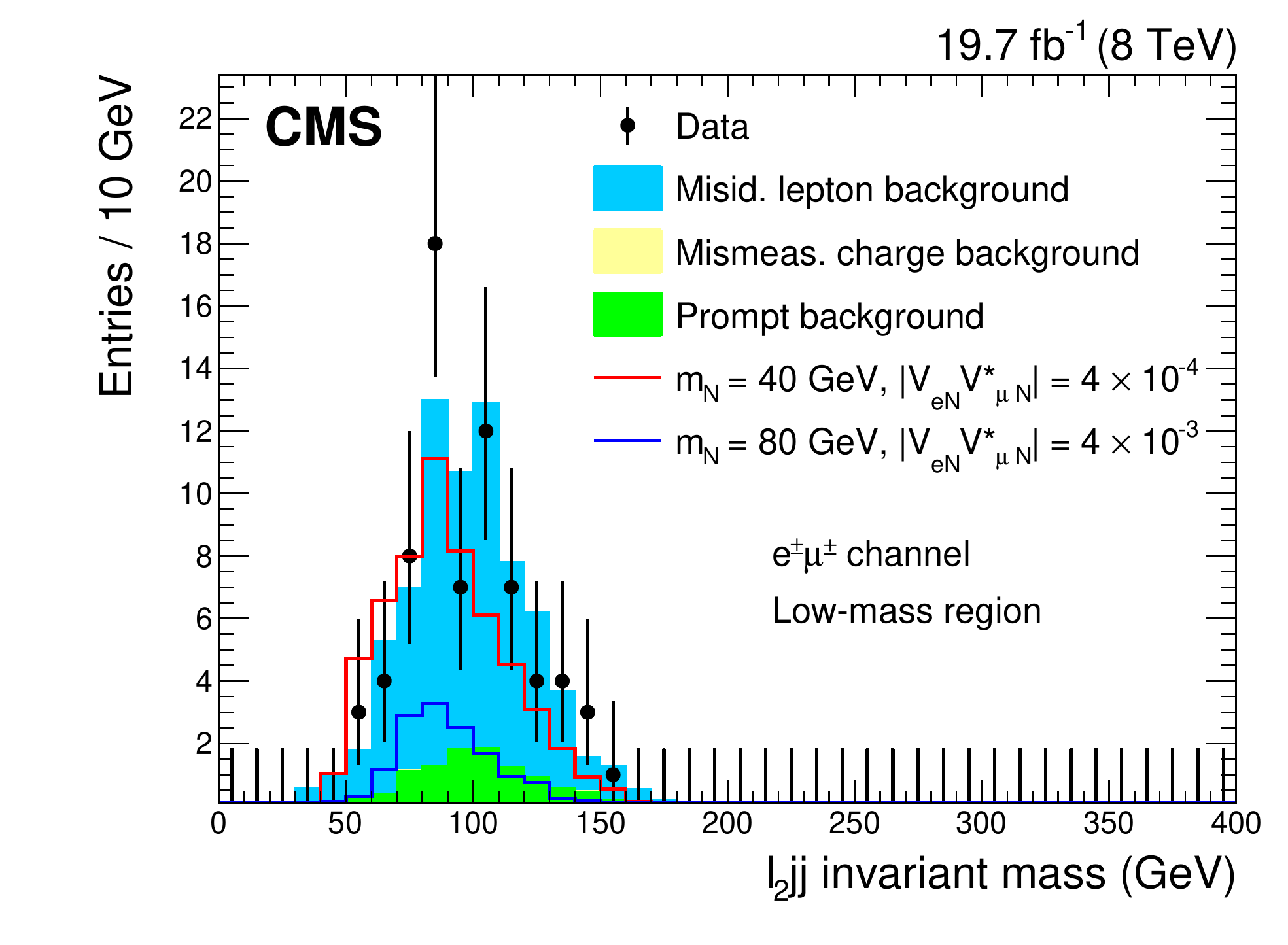}
\includegraphics[width=0.45\textwidth]{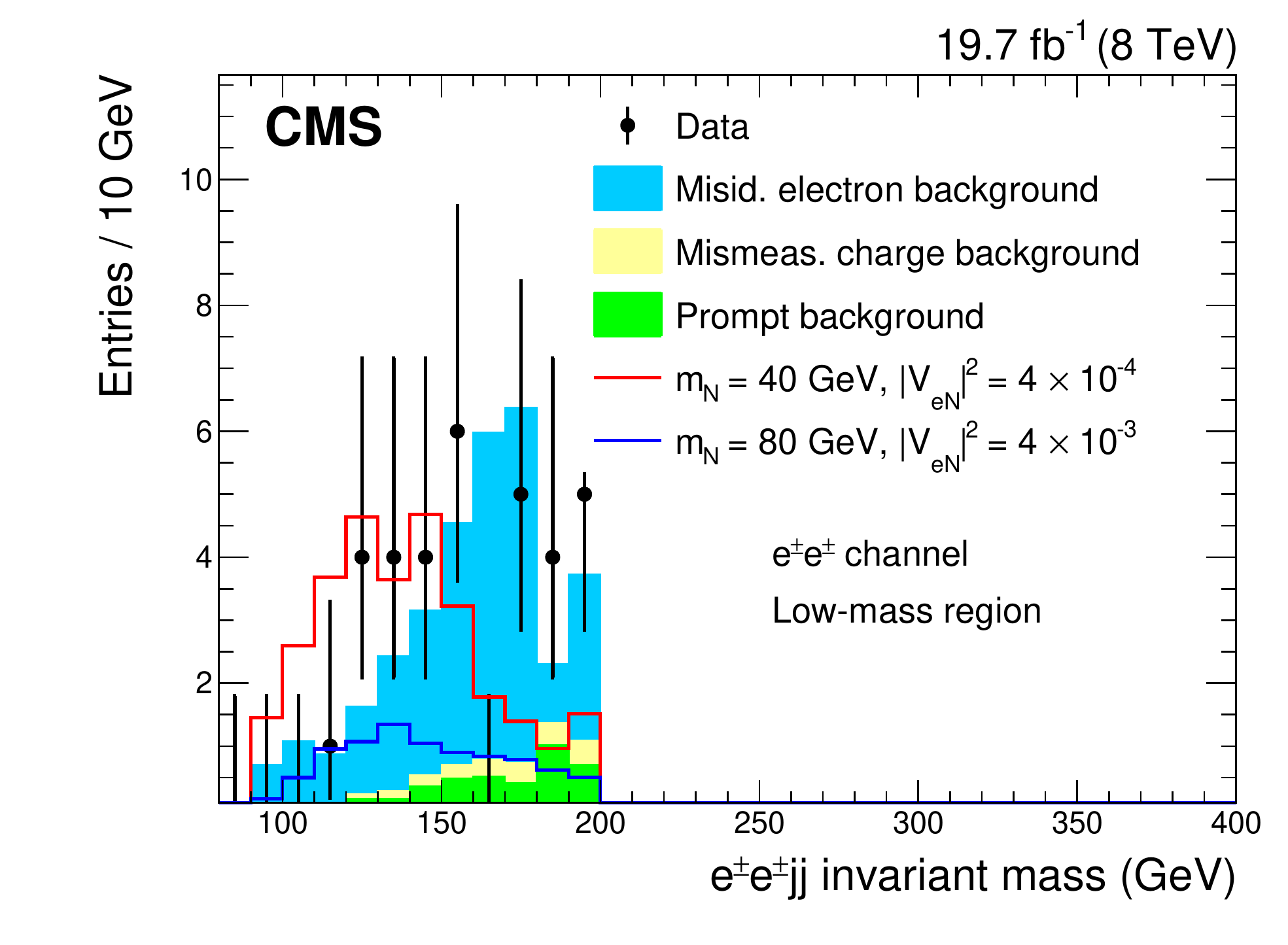} \includegraphics[width=0.45\textwidth]{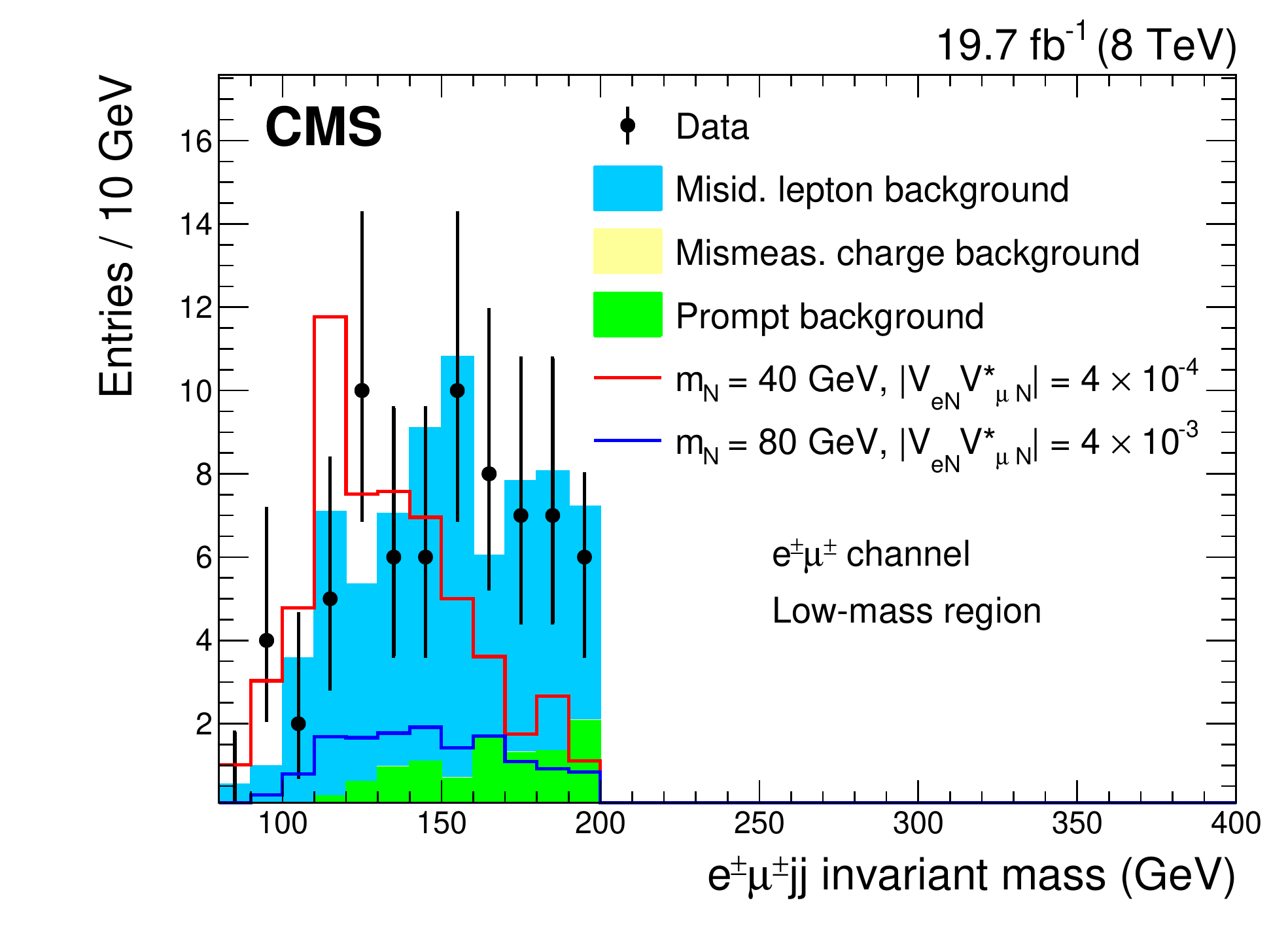}
\includegraphics[width=0.45\textwidth]{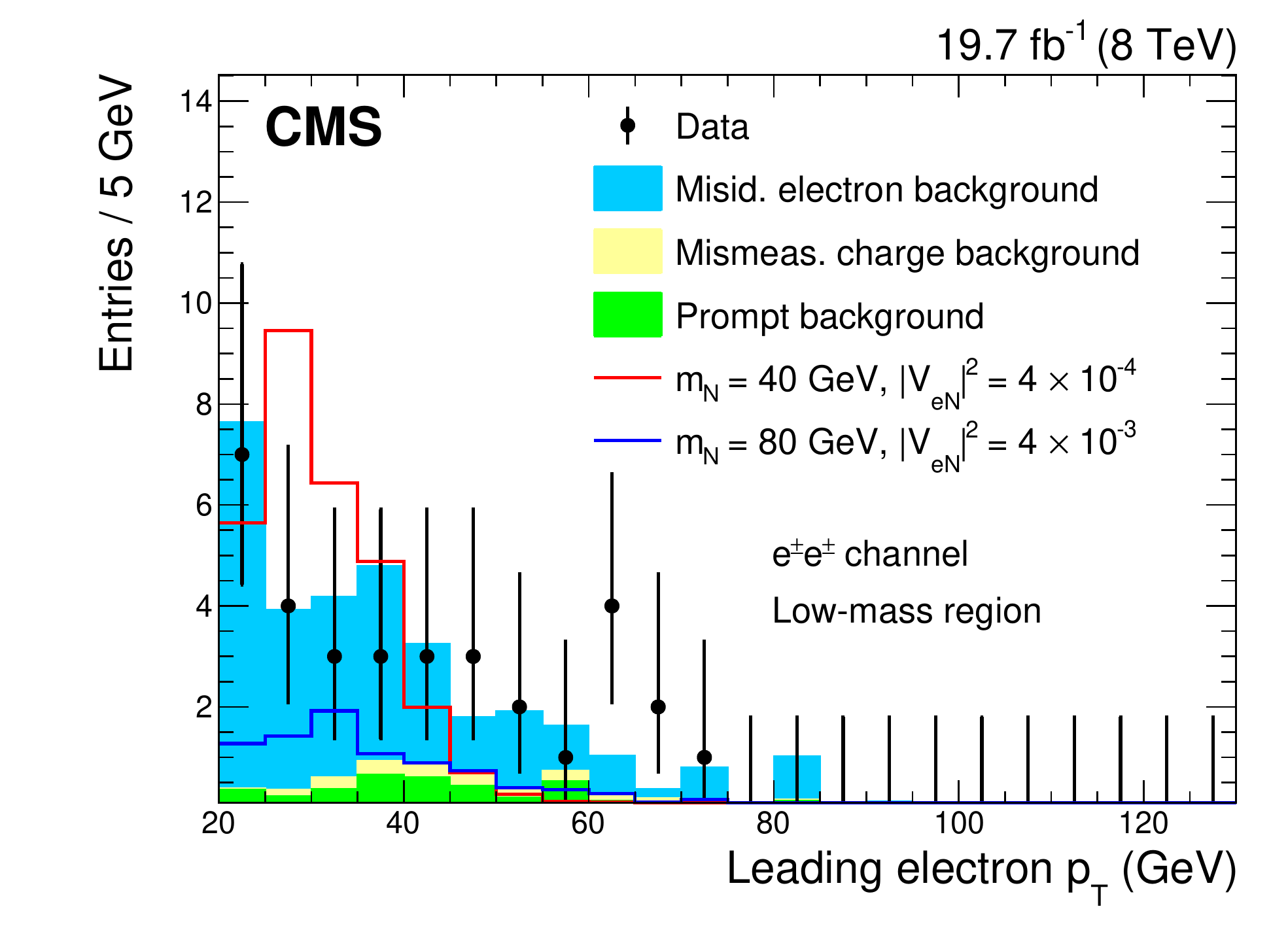} \includegraphics[width=0.45\textwidth]{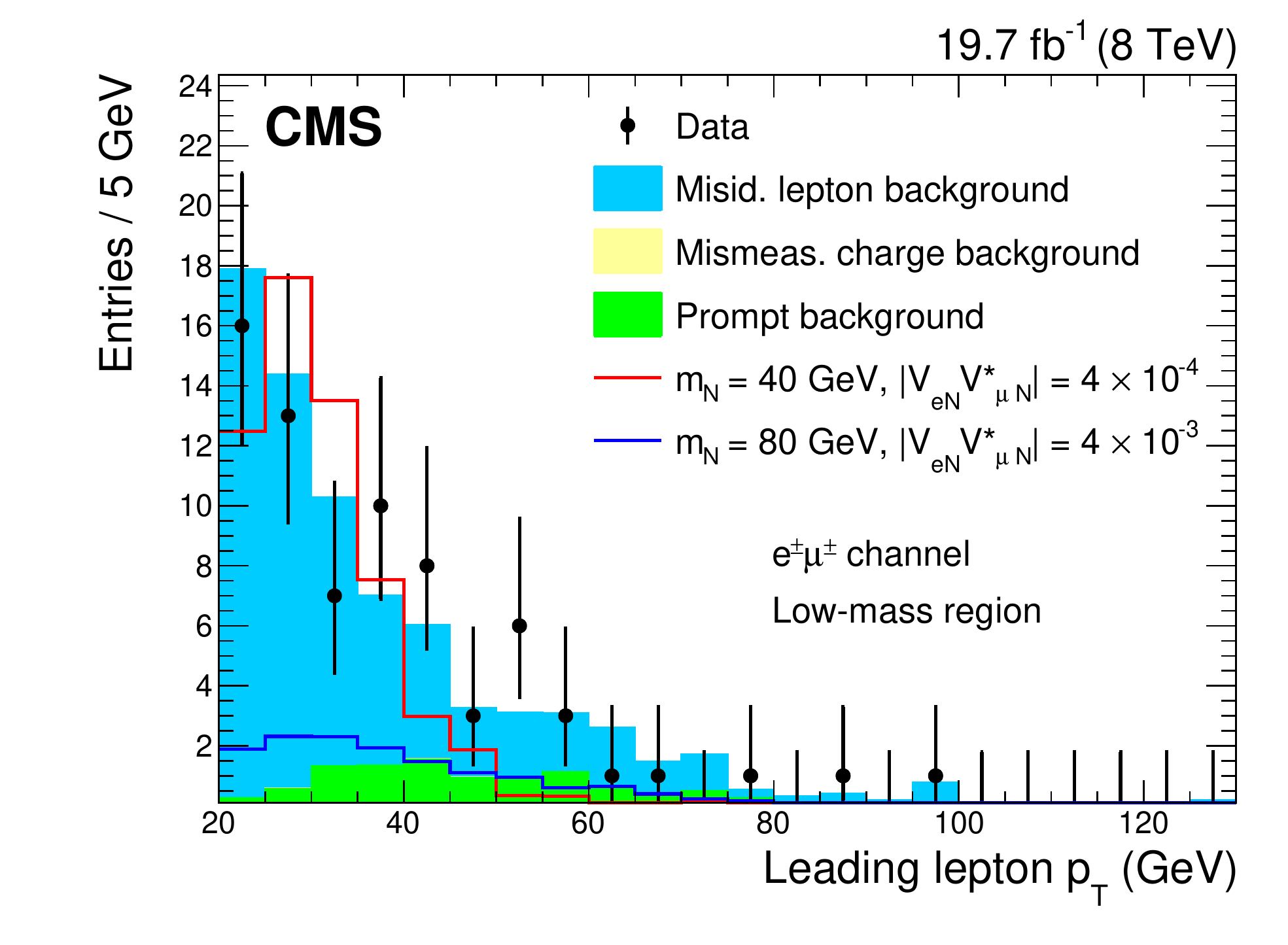}
\caption{Kinematic distributions for the low-mass region after all selection cuts are applied 
except for the final optimization requirement:
dielectron channel (left), electron-muon channel (right).
The plots show the data, backgrounds, and two choices for the heavy Majorana neutrino signal.
}
\label{fig:mass_low}
\end{center}
\end{figure}

\begin{figure}[tbph]
\begin{center}
\includegraphics[width=0.45\textwidth]{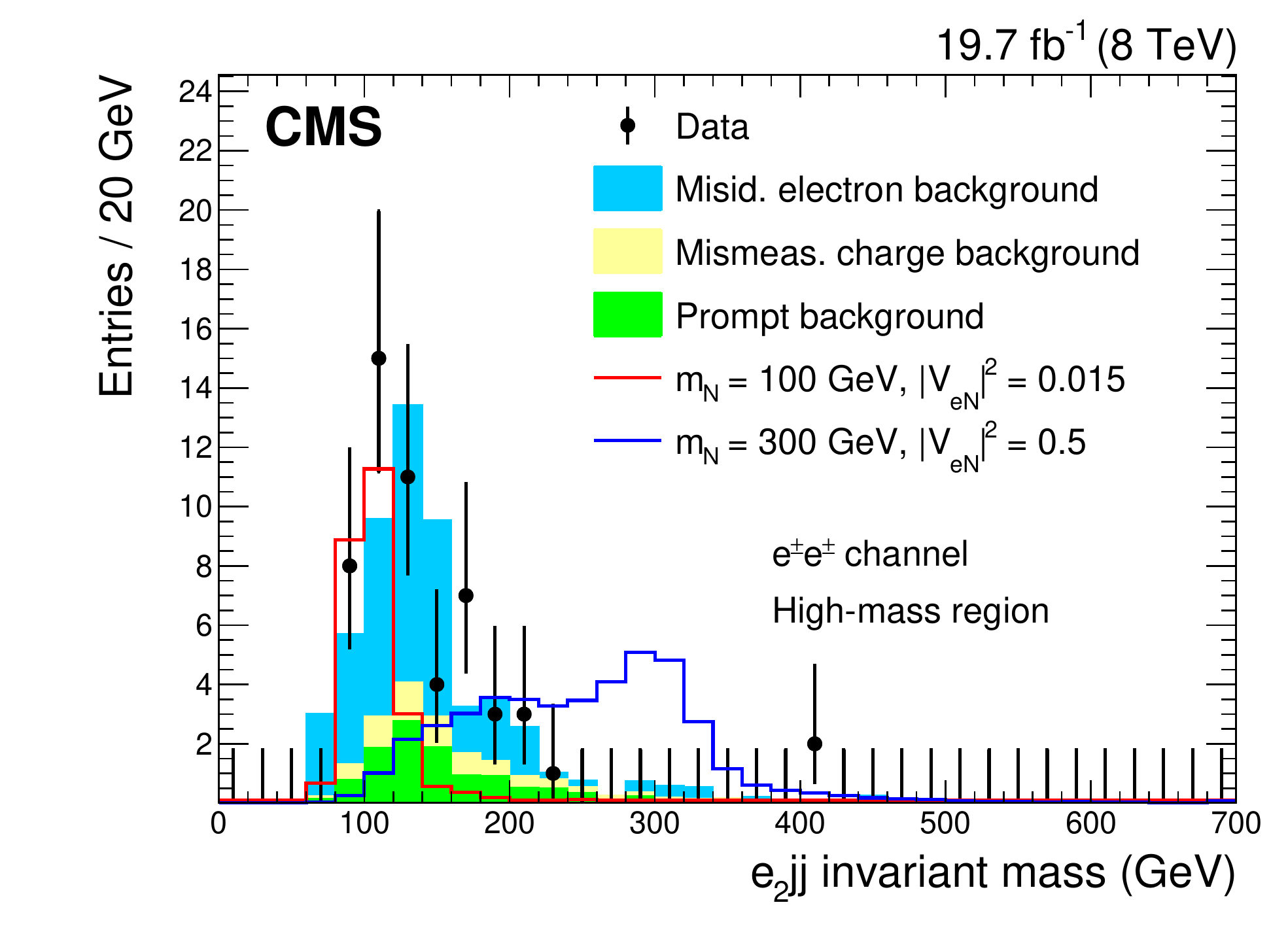} \includegraphics[width=0.45\textwidth]{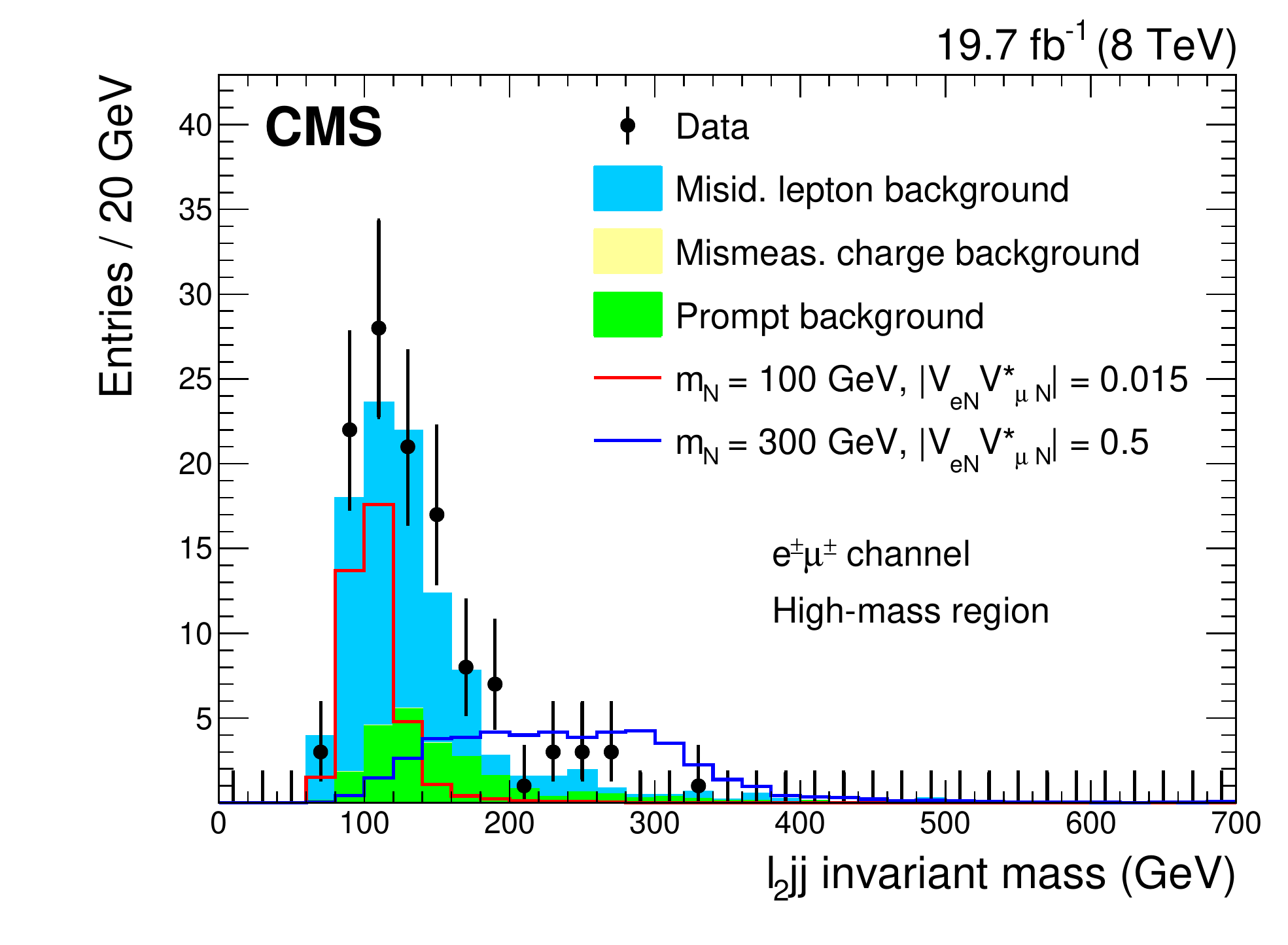}
\includegraphics[width=0.45\textwidth]{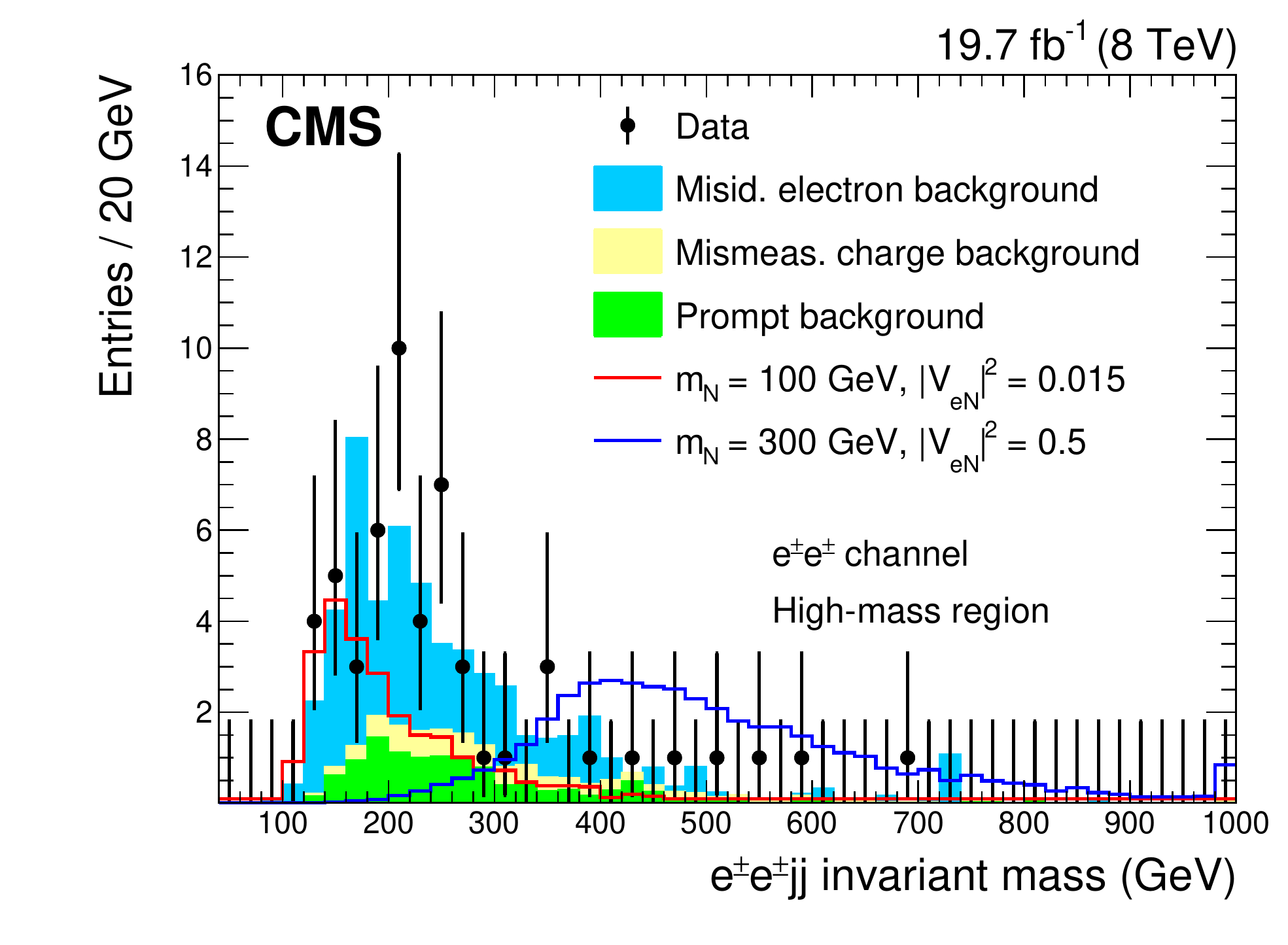} \includegraphics[width=0.45\textwidth]{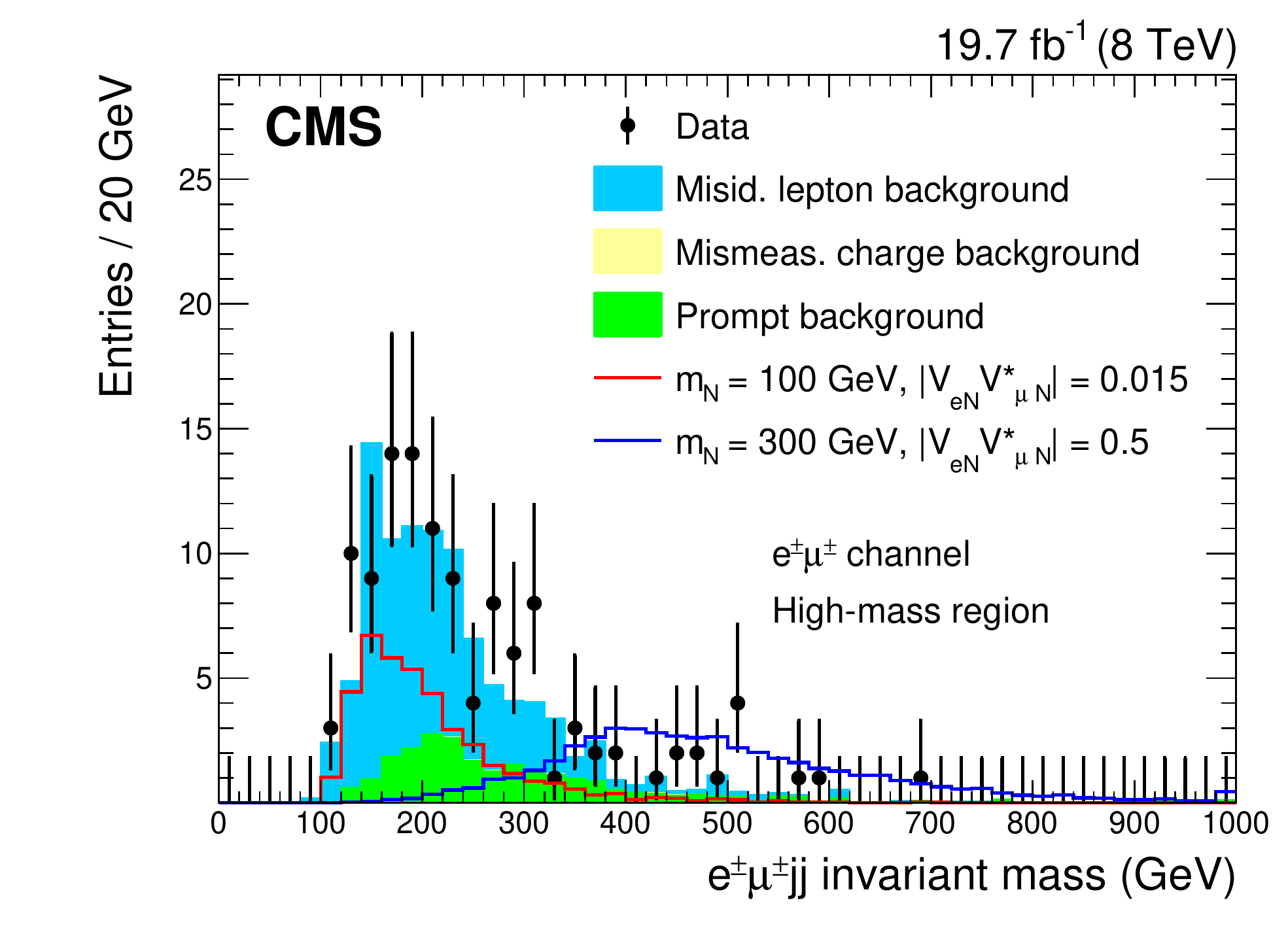}
\includegraphics[width=0.45\textwidth]{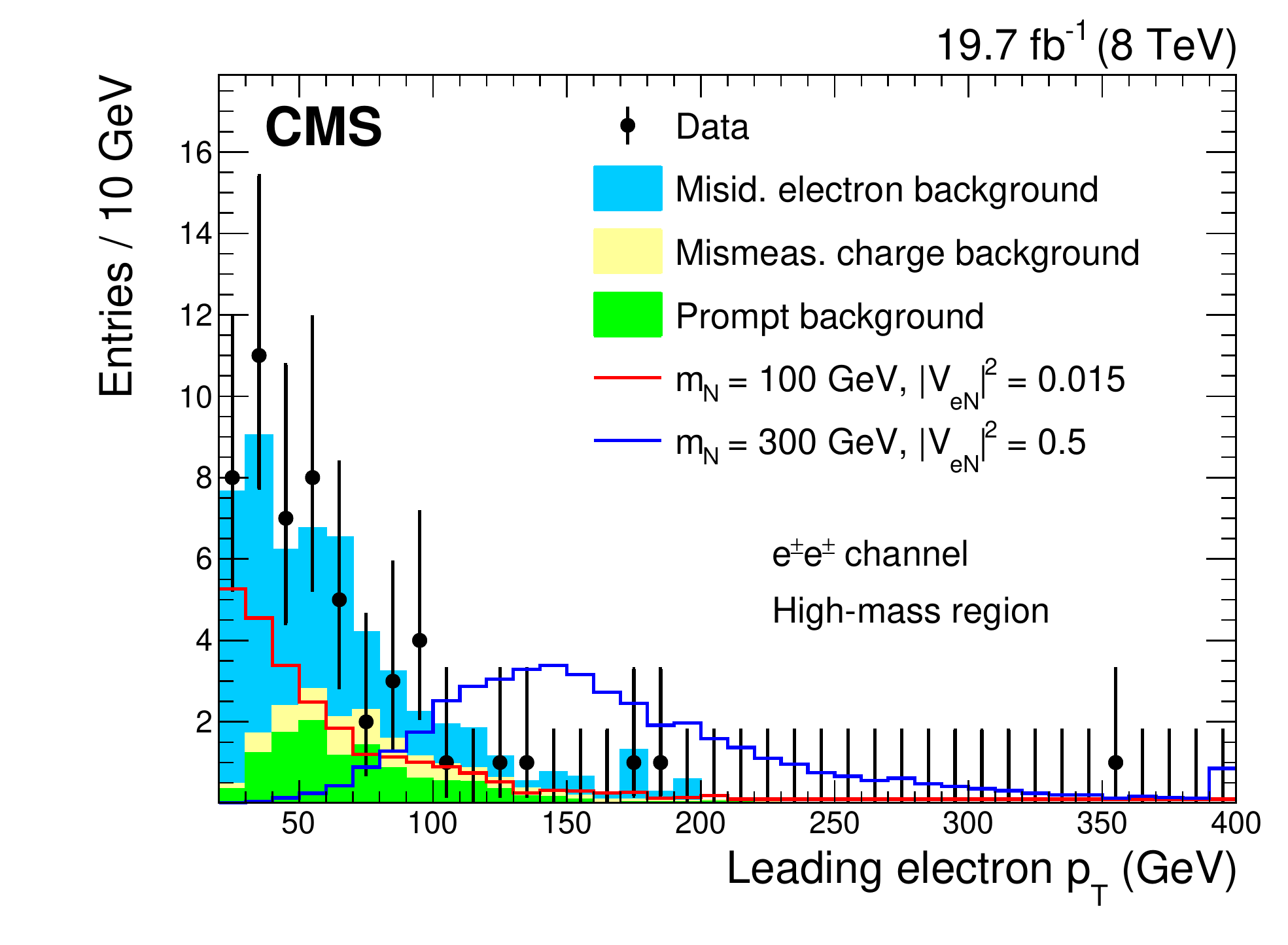} \includegraphics[width=0.45\textwidth]{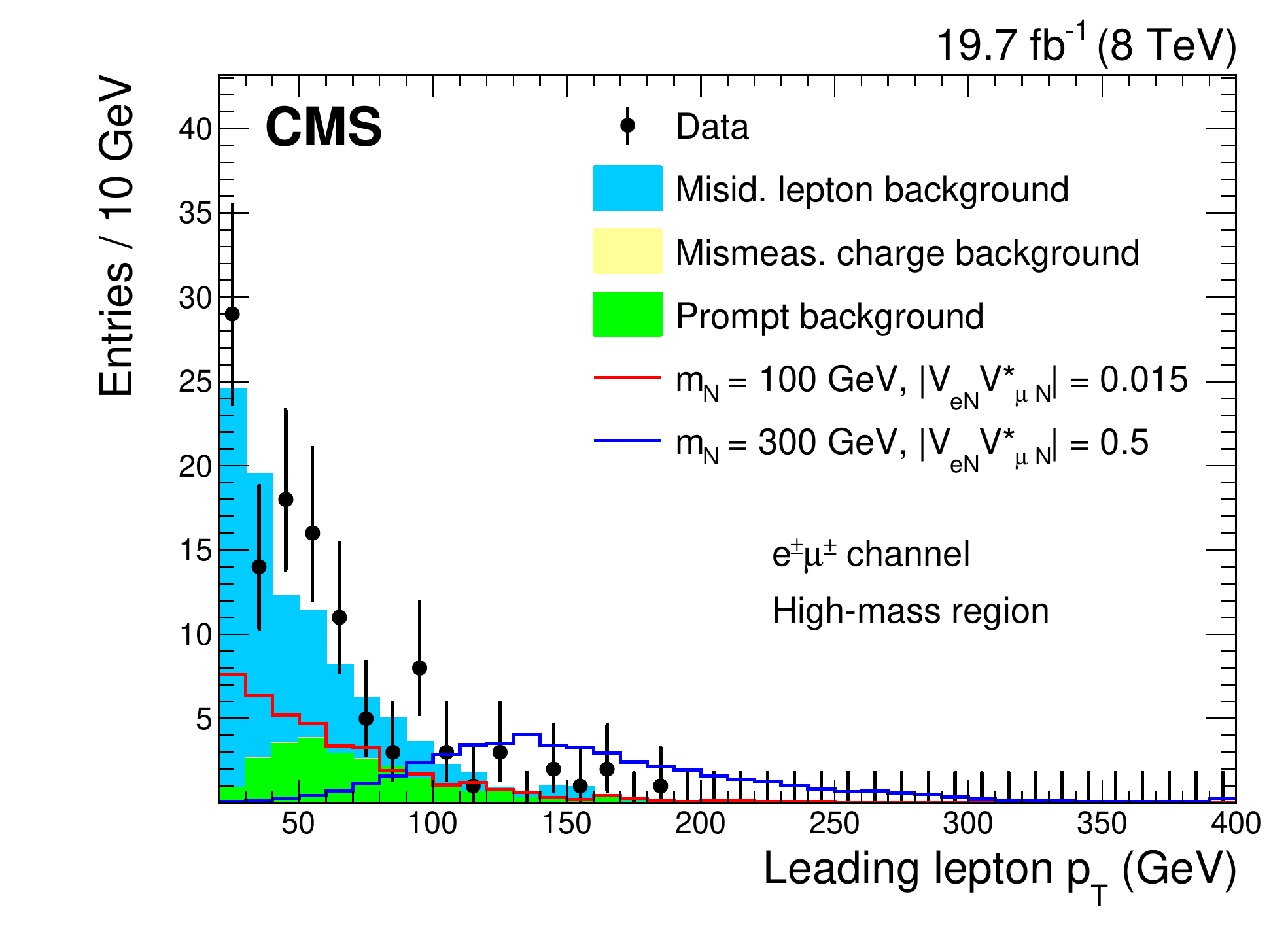}
\caption{Kinematic distributions for the high-mass region after all selection cuts are applied except for the final optimization requirement: 
dielectron channel (left), electron-muon channel (right).
The plots show the data, backgrounds, and two choices for the heavy Majorana neutrino signal.
}
\label{fig:mass_high}
\end{center}
\end{figure}

The data yields and background estimates after the application of all selection cuts, except the final optimization 
cuts are shown in Table~\ref{table:yields_before_opt}. The data yields are in good agreement with the estimated backgrounds. 
Kinematic distributions also show good agreement between data and backgrounds. Figures~\ref{fig:mass_low} and \ref{fig:mass_high}
show some of the kinematic distributions: the invariant mass of the trailing $\pt$ lepton and the two 
selected jets; the invariant mass of the two leptons and the two selected jets; and the leading lepton $\pt$. 
The background predictions from prompt same-sign leptons and misidentified leptons
are shown along with the total background estimate and the number of events observed in data. 
The uncertainties shown are the statistical and systematic components, respectively. 
In Fig.~\ref{fig:mass_low}, the $m(\Pe^\pm \Pe^\pm \mathrm{jj})$ signal distribution is not peaked at $m_W$ because of 
the kinematic requirements imposed.
\begin{table*}[htb]
\topcaption{Observed event yields and estimated backgrounds after the application of all selection, except for the final optimization.
The background predictions from prompt same-sign leptons (Prompt bkgd.), misidentified leptons (Misid. bkgd.), 
mismeasured charge (Charge mismeas. bkgd.) and the total background (Total bkgd.) are shown together with
the number of events observed in data. The uncertainties shown are the statistical and systematic components, respectively. 
}
\label{table:yields_before_opt}
\begin{center}
\resizebox{\columnwidth}{!}{
\begin{tabular}{cccccc}
\hline 
Channel / Region &  Prompt bkgd. & Misid. bkgd. & Charge mismeas. bkgd. & Total bkgd. & $N_{\mathrm{obs}}$ \\
\hline
$\Pe\Pe$ / Low-mass & \phantom{0}4.0 $\pm$ 0.4  $\pm$ 0.8  & 26.7  $\pm$ 3.2 $\pm$ 10.7 & 2.00 $\pm$ 0.03 $\pm$ 0.24 & 32.6 $\pm$ 3.2 $\pm$ 10.7 & 33 \\
$\Pe\Pe$ / High-mass & 10.8 $\pm$ 0.7  $\pm$ 2.2 & 36.9  $\pm$ 3.6 $\pm$ 14.8 & 6.99 $\pm$ 0.09 $\pm$ 0.84 & 55.4 $\pm$ 3.6 $\pm$ 14.8 & 54 \\
 $\Pe\mu$ / Low-mass & 10.4 $\pm$ 0.7  $\pm$ 2.1  & 63.4  $\pm$ 4.1 $\pm$ 21.5 & 0.07 $\pm$ 0.01 $\pm$ 0.01 & 73.9 $\pm$ 4.1 $\pm$ 21.6 & 71 \\
 $\Pe\mu$ / High-mass & 24.1 $\pm$ 1.1  $\pm$ 4.8 & 75.6  $\pm$ 4.3 $\pm$ 25.7 & 0.24 $\pm$ 0.01 $\pm$ 0.01 & 99.8 $\pm$ 4.5 $\pm$ 25.8 & 117 \\
\hline
\end{tabular}
}
\end{center}
\end{table*}
\begin{table*}[p]
\topcaption{Dielectron and electron-muon channel results after final optimization.
The background predictions from  prompt same-sign leptons, misidentified leptons, 
and mismeasured charge are shown along with the total background estimate and the number of events observed in data. 
The uncertainties shown are the statistical and systematic components, respectively. 
}
\label{table:yields}
\begin{center}
\resizebox{\columnwidth}{!}{
\begin{tabular}{cccccc}
\hline 
$\mN$ &  \multirow{2}{*}{Prompt bkgd.} & \multirow{2}{*}{Misid. bkgd.} & \multirow{2}{*}{Charge mismeas. bkgd.} & \multirow{2}{*}{Total bkgd.} & \multirow{2}{*}{$N_{\mathrm{obs}}$}  \\
(\GeVns)  &  & & & & \\
\hline\
$\Pe\Pe$ channel: &  & & & & \\
40-80  & 0.8 $\pm$ 0.2  $\pm$ 0.1    & 7.5  $\pm$ 2.0 $\pm$ 3.0 & 0.27 $\pm$ 0.01 $\pm$ 0.03 & 8.6  $\pm$ 2.0 $\pm$ 3.0 &  11 \\
90      & 2.8 $\pm$ 0.3  $\pm$ 0.3    & 13.4 $\pm$ 2.2 $\pm$ 5.4  & 1.68 $\pm$ 0.04 $\pm$ 0.20 & 17.8 $\pm$ 2.2 $\pm$ 5.4 &  23 \\
100     & 2.6 $\pm$ 0.3  $\pm$ 0.3    & 11.0 $\pm$ 2.1 $\pm$ 4.5  & 1.60 $\pm$ 0.04 $\pm$ 0.19 & 15.3 $\pm$ 2.1 $\pm$ 4.5 &  23 \\
125     & 3.3 $\pm$ 0.4  $\pm$ 0.4    & 6.1  $\pm$ 1.3 $\pm$ 2.4  & 1.72 $\pm$ 0.04 $\pm$ 0.21 & 11.1 $\pm$ 1.3 $\pm$ 2.5 &  11 \\
150     & 3.3 $\pm$ 0.4  $\pm$ 0.4    & 4.7  $\pm$ 1.1 $\pm$ 1.9  & 1.93 $\pm$ 0.05 $\pm$ 0.23 & 9.9 $\pm$ 1.2 $\pm$ 1.9 &  7 \\
175     & 2.0 $\pm$ 0.3  $\pm$ 0.3    & 0.9  $\pm$ 0.5 $\pm$ 0.4  & 1.10 $\pm$ 0.04 $\pm$ 0.13 & 4.0  $\pm$ 0.6 $\pm$ 0.5 &  3 \\
200     & 1.3 $\pm$ 0.2  $\pm$ 0.2    & 2.0  $\pm$ 1.3 $\pm$ 0.8  & 1.02 $\pm$ 0.04 $\pm$ 0.12 & 4.3  $\pm$ 1.3 $\pm$ 0.8 &  3 \\
250     & 1.1 $\pm$ 0.2  $\pm$ 0.2    & 1.8  $\pm$ 1.4 $\pm$ 0.8  & 0.84 $\pm$ 0.04 $\pm$ 0.10 & 3.8  $\pm$ 1.4 $\pm$ 0.7 &  4 \\
300     & 0.8 $\pm$ 0.2  $\pm$ 0.1    & 1.2  $\pm$ 1.3 $\pm$ 0.5  & 0.66 $\pm$ 0.04 $\pm$ 0.08 & 2.6  $\pm$ 1.3 $\pm$ 0.5 &  4 \\
350     & 0.6 $\pm$ 0.2  $\pm$ 0.1    & 1.2  $\pm$ 1.3 $\pm$ 0.5  & 0.59 $\pm$ 0.04 $\pm$ 0.07 & 2.4  $\pm$ 1.3 $\pm$ 0.5 &  4 \\
400     & 0.6 $\pm$ 0.2  $\pm$ 0.1    & 1.2  $\pm$ 1.3 $\pm$ 0.5  & 0.59 $\pm$ 0.04 $\pm$ 0.07 & 2.4  $\pm$ 1.3 $\pm$ 0.5 &  4 \\
500     & 0.6 $\pm$ 0.2  $\pm$ 0.1    & 1.2  $\pm$ 1.3 $\pm$ 0.5  & 0.59 $\pm$ 0.04 $\pm$ 0.07 & 2.4  $\pm$ 1.3 $\pm$ 0.5 &  4 \\
 $\Pe\mu$ channel: &  & & & & \\
40-70   & 3.1 $\pm$ 0.3  $\pm$ 0.5    & 30.6  $\pm$ 3.0 $\pm$ 10.4  &---   & 33.7  $\pm$ 3.0 $\pm$ 10.4 &  33 \\
80      & 8.1 $\pm$ 0.6  $\pm$ 1.2    & 17.2  $\pm$ 1.8 $\pm$ 5.9   &---  & 25.3  $\pm$ 1.9 $\pm$ 6.0 &  29\\
90      & 6.6 $\pm$ 0.6  $\pm$ 1.0    & 13.4 $\pm$ 1.4 $\pm$ 4.6   &---     & 20.1 $\pm$ 1.6 $\pm$ 4.6 &  25 \\
100     & 6.7 $\pm$ 0.6  $\pm$ 1.1    & 8.1 $\pm$ 1.0 $\pm$ 2.7    &---     & 14.8  $\pm$ 1.2 $\pm$ 2.9 &  20 \\
125     & 7.2 $\pm$ 0.6  $\pm$ 1.2    & 5.1  $\pm$ 0.9 $\pm$ 1.7   &---     & 12.3 $\pm$ 1.1 $\pm$ 1.9 &  17 \\
150     & 8.2 $\pm$ 0.6  $\pm$ 1.2    & 5.6  $\pm$ 0.9 $\pm$ 1.9   &---    & 13.8 $\pm$ 1.1 $\pm$ 2.3 &  16 \\
175     & 5.6 $\pm$ 0.5  $\pm$ 0.8    & 3.6  $\pm$ 0.7 $\pm$ 1.2   &---     & 9.3  $\pm$ 0.9 $\pm$ 1.5 &  11 \\
200     & 3.7 $\pm$ 0.4  $\pm$ 0.6    & 2.5  $\pm$ 0.6 $\pm$ 0.8   &---    &6.2  $\pm$ 0.7  $\pm$ 1.0 &  7 \\
250     & 3.1 $\pm$ 0.4  $\pm$ 0.5    & 1.5  $\pm$ 0.5 $\pm$ 0.5   &---     & 4.7  $\pm$ 0.6 $\pm$ 0.6 &  7 \\
300     & 1.4 $\pm$ 0.2  $\pm$ 0.2    & 0.7  $\pm$ 0.3 $\pm$ 0.2   &---    &2.2  $\pm$ 0.4 $\pm$ 0.3 &  4 \\
350     & 0.9 $\pm$ 0.2  $\pm$ 0.1    & 0.7  $\pm$ 0.3 $\pm$ 0.2   &---    &1.6  $\pm$ 0.4 $\pm$ 0.3 &  4 \\
400     & 0.8 $\pm$ 0.2  $\pm$ 0.1    & 0.7  $\pm$ 0.3 $\pm$ 0.2   &---    &1.6  $\pm$ 0.4 $\pm$ 0.3 &  4 \\
500     & 0.8 $\pm$ 0.2  $\pm$ 0.1    & 0.7  $\pm$ 0.3 $\pm$ 0.2   &---    &1.6  $\pm$ 0.4 $\pm$ 0.3 &  4 \\
\hline
\end{tabular}
}
\end{center}
\end{table*}

After applying all the final optimized selections, the background estimates and numbers of observed events are shown in 
Table~\ref{table:yields}. The expected signal depends on $\mN$ and the mixing 
$\abs{ \VeN }^2$ or $\abs{\VeN \VmNst}^2 / ( |\VeN|^2 + |\VmN|^2 )$. For the dielectron (electron-muon) channel, the expected number of signal events for 
$\mN = 50\GeV$ and $\abs{\VeN}^2 = 1 \times 10^{-3}~(\abs{\VeN \VmNst}^2 / ( |\VeN|^2 + |\VmN|^2 ) = 1 \times 10^{-3})$ is 74 (256). For 
$\mN = 100\GeV$ and $\abs{\VeN}^2 = 1 \times 10^{-3}~(\abs{\VeN \VmNst}^2 / ( |\VeN|^2 + |\VmN|^2 ) = 1 \times 10^{-3})$ it is 1.8 (3.6) events, while for
$\mN = 500\GeV$ and $\abs{\VeN}^2 = 1 (\abs{\VeN \VmNst}^2 / ( |\VeN|^2 + |\VmN|^2 ) = 1)$ it is 9.2 (13.8) events.

No  significant excess in the data compared to the backgrounds predicted from the SM is seen and 95\% 
confidence level (CL) exclusion limits are set on the Majorana neutrino mixing element and cross section times branching 
fraction for $\Pp\Pp \rightarrow \N \ell^{\pm} \rightarrow \ell^{\pm} \ell^{(\prime)\pm} \PQq\PAQq^\prime$ 
as a function of $\mN$. The limits are obtained using the CLs method~\cite{cls,cls2,lhc_limit} based on the event yields in 
Table~\ref{table:yields}. Poisson distributions are used for the signal and log-normal 
distributions for the nuisance parameters. Limits are also set on $| \VeN |^2$ and $\abs{\VeN \VmNst}^2 / ( |\VeN|^2 + |\VmN|^2 )$ as a 
function of $\mN$.
 
\begin{figure}[hbtp]\begin{center}
  \includegraphics[width=0.65\textwidth]{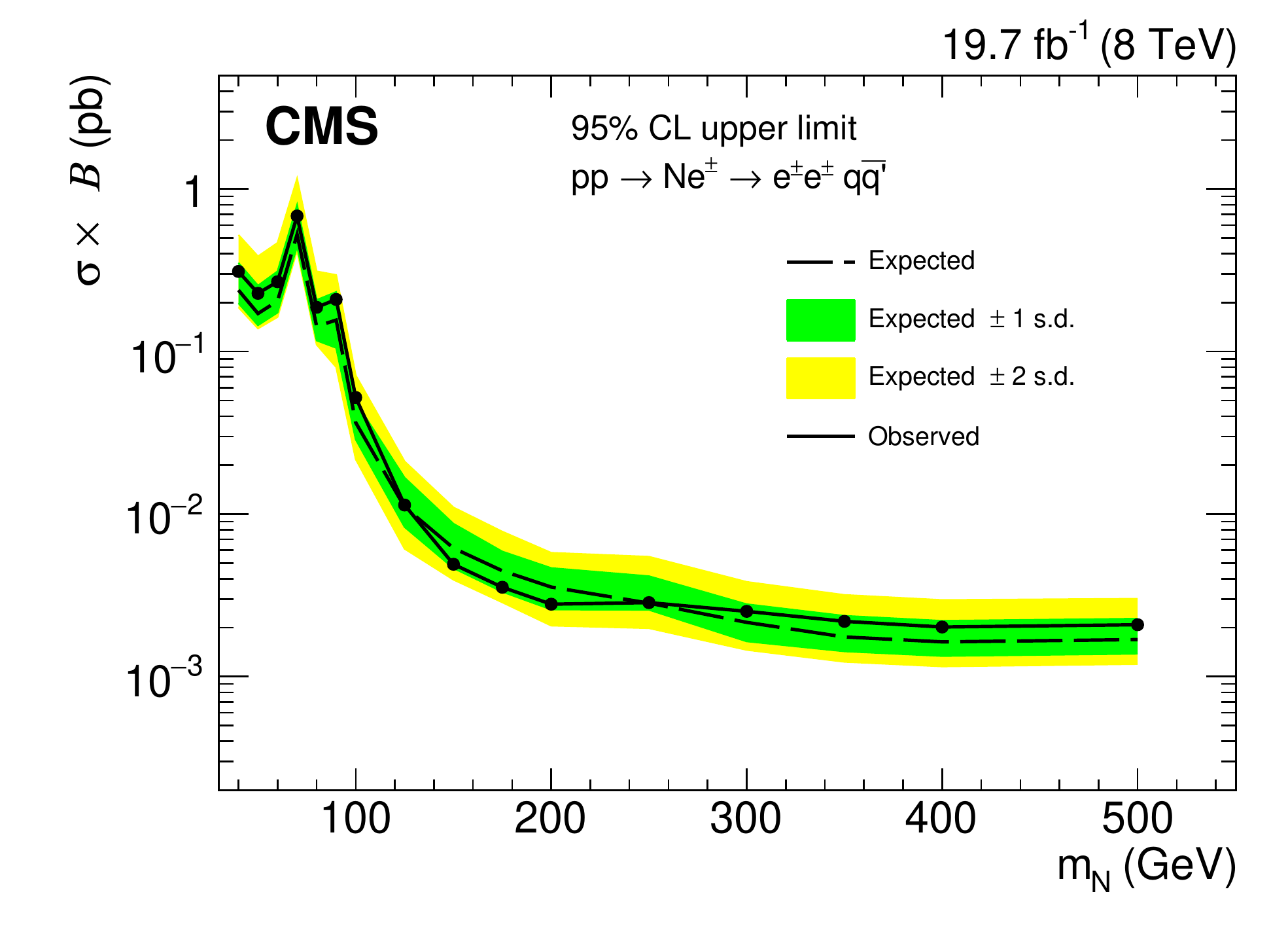}
  \includegraphics[width=0.65\textwidth]{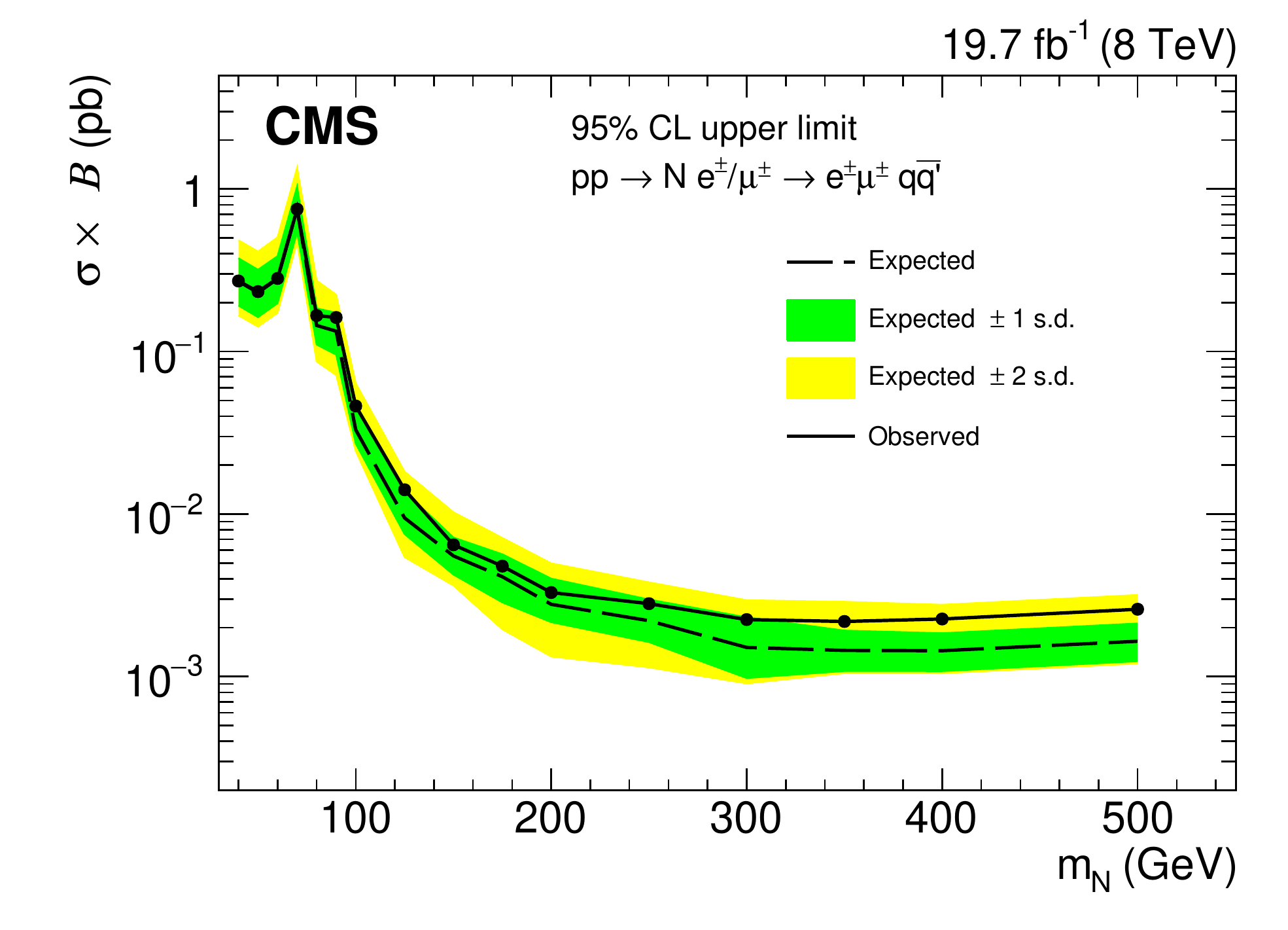}
  \caption{ Exclusion region at 95\% CL in the cross section times branching fraction for
  $\sigma(\Pp\Pp \rightarrow \N \Pe^{\pm} \rightarrow \Pe^\pm \Pe^\pm \PQq \PAQq^\prime)$
  (top) and
  $\sigma(\Pp\Pp \rightarrow \N \; \Pe^{\pm} / \mu^\pm \rightarrow  \Pe^\pm \mu^\pm \PQq \PAQq^\prime)$
  (bottom)
  as a function of \mN. 
  The dashed curve is the expected upper limit, with one and two
  standard-deviation bands shown in dark green and light yellow, respectively. The solid black curve is the
  observed upper limit.}
\label{fig:excl_xs}
\end{center}
\end{figure}

\begin{figure}[hbtp]\begin{center}
  \includegraphics[width=0.65\textwidth]{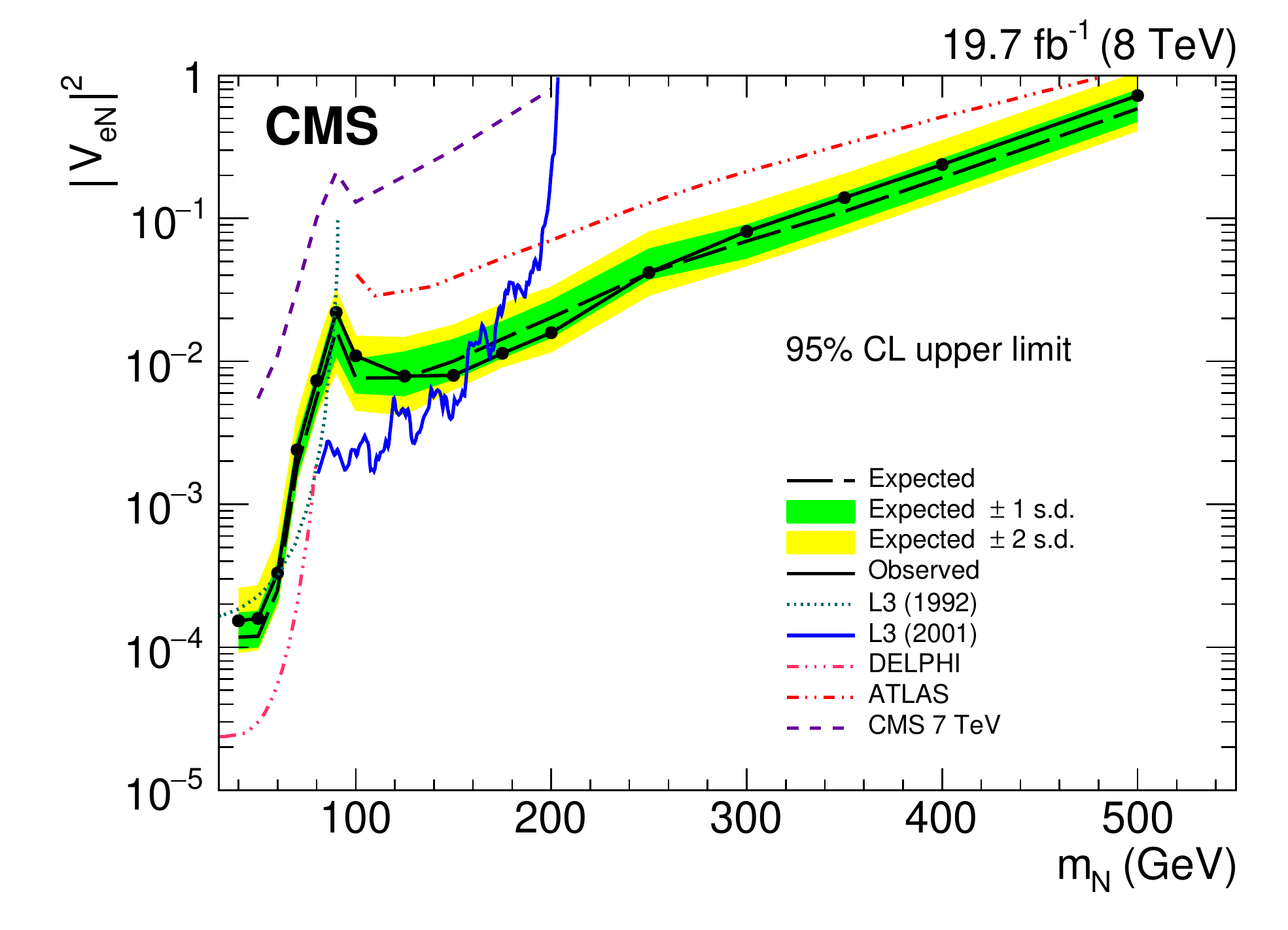}\label{fig:excl}
  \includegraphics[width=0.65\textwidth]{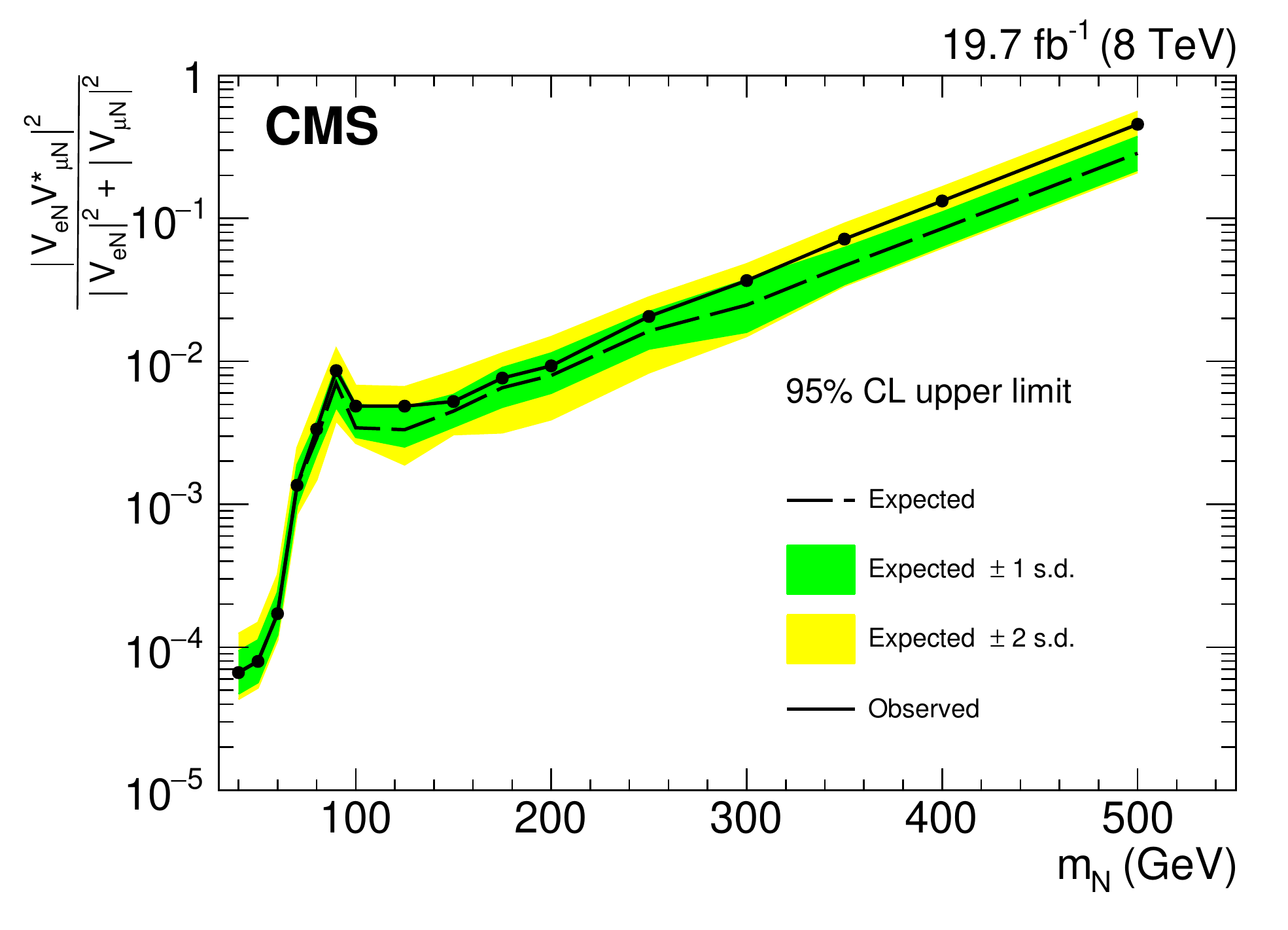}\label{fig:excl_em}
  \caption{Exclusion region at 95\% CL in the $\abs{\VeN}^2$ vs. $\mN$ plane (top) and 
  $\abs{ \VeN }^2$ and $\abs{\VeN \VmNst}^2 / ( |\VeN|^2 + |\VmN|^2 )$  vs. $\mN$ plane (bottom).
  The dashed black curve is the expected upper limit, with one and two
  standard-deviation bands shown in dark green and light yellow, respectively. The solid black curve is the
  observed upper limit. Also shown are the upper limits from other direct searches: L3~\cite{l3, l3_2001}, DELPHI~\cite{delphi}, ATLAS~\cite{ATLAS_NR_2012},
  and the upper limits from the CMS $\sqrt{s} = 7\TeV$ (2011) data~\cite{CMS_NR_2011}.
  }
\label{fig:excl_mix}
\end{center}
\end{figure}

The 95\% CL limits on the cross section times branching fractions,
$\Pp\Pp \rightarrow \N \ell^{\pm} \rightarrow \ell^{\pm} \ell^{(\prime)\pm} \PQq\PAQq^\prime$, as a function of $\mN$, 
are shown in Fig.~\ref{fig:excl_xs}.
The limits on the absolute values of the mixing elements $\abs{ \VeN }^2$ and $\abs{\VeN \VmNst}^2 / ( |\VeN|^2 + |\VmN|^2 )$ 
are shown in Fig.~\ref{fig:excl_mix}, also as a function of $\mN$. 
The mass range below $\mN = 40\GeV$ is not considered because of 
the very low selection efficiency for the signal in this mass region. The behaviour of the limits around $\mN = 80\GeV$ 
is caused by the fact that as the heavy Majorana neutrino gets close to the \PW boson mass from below or above, the 
lepton produced together with the \N or the lepton from the \N decay has very low $\pt$, respectively. 

A significant increase in sensitivity on the limits for $\abs{ \VeN }^2$ has been achieved over the 
previous limits set by CMS with the 7\TeV data set~\cite{CMS_NR_2011}. These limits are the most restrictive direct limits for heavy Majorana 
neutrino masses above 200\GeV.  The limits on $\abs{ \VeN }^2$ and $\abs{\VeN \VmNst}^2 / ( |\VeN|^2 + |\VmN|^2 )$  presented here are the first direct limits on this quantity for $\mN > 40\GeV$. 

\begin{figure}[hbtp]\begin{center}
  \includegraphics[width=0.65\textwidth]{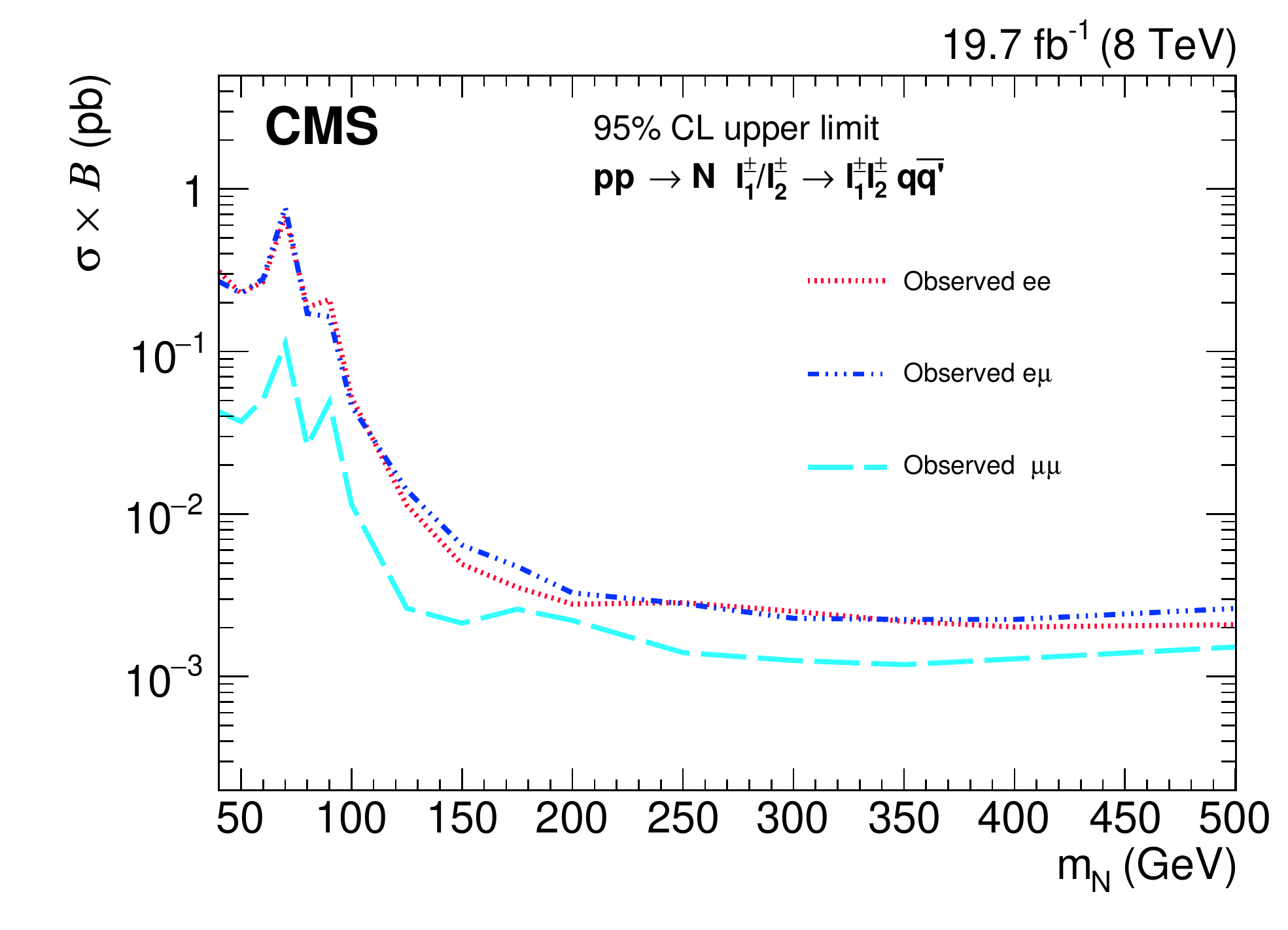}
  \caption{Comparison of observed exclusion regions at 95\% CL in the cross section times branching fraction as a function of $\mN$ for
   $\Pp\Pp \rightarrow \N \Pe^{\pm}\rightarrow \Pe^{\pm} \Pe^{\pm}\PQq\PAQq^\prime$,
   $\Pp\Pp \rightarrow \N \mu^{\pm} \rightarrow \mu^{\pm} \mu^{\pm}\PQq\PAQq^\prime$, and
  $\Pp\Pp \rightarrow \N \; \Pe^{\pm} / \mu^{\pm} \rightarrow \Pe^{\pm} \mu^{\pm}\PQq\PAQq^\prime$. 
  The result for the dimuon channel is from Ref.~\cite{CMS_NR_mu_2012}.
  }
\label{fig:Figure_006}
\end{center}
\end{figure}

\begin{figure}[hbtp]\begin{center}
  \includegraphics[width=0.65\textwidth]{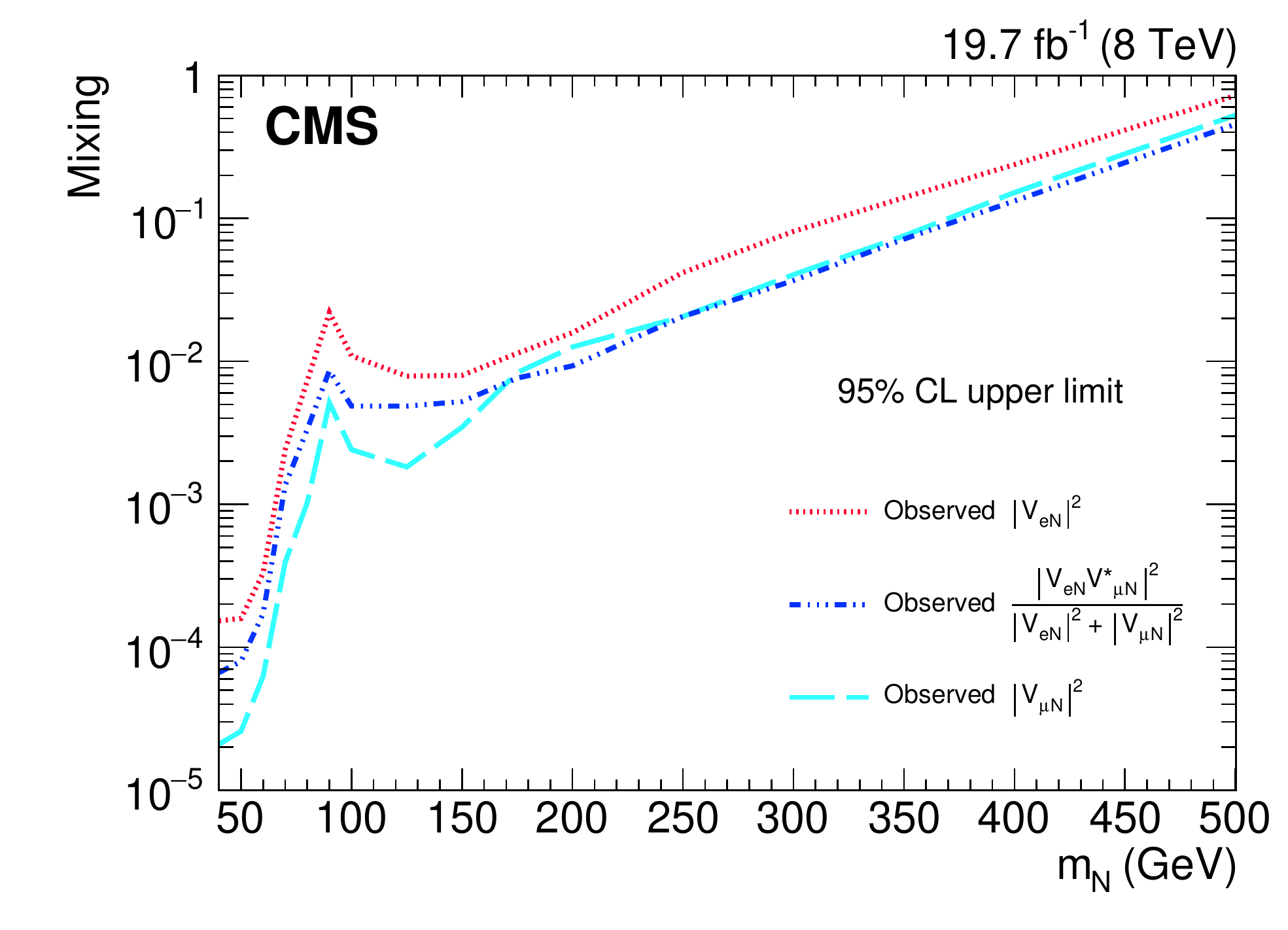}
  \caption{Comparison of observed exclusion regions at 95\% CL in
  $\abs{\VmN}^2$, $\abs{\VeN}^2$, and $\abs{\VeN \VmNst}^2 / ( |\VeN|^2 + |\VmN|^2 )$. The CMS result for  $\abs{\VmN}^2$ is 
  from Ref.~\cite{CMS_NR_mu_2012}.
  }
\label{fig:Figure_007}
\end{center}
\end{figure}

The observed limits on the cross section times branching fraction for the dielectron and electron-muon channels are shown 
in Fig.~\ref{fig:Figure_006} along with the corresponding CMS limit for the dimuon channel obtained at $\sqrt{s} = 8\TeV$ 
reported in Ref.~\cite{CMS_NR_mu_2012}. A similar comparison of the limits on the mixing elements is shown in Fig.~\ref{fig:Figure_007}.

\section{Summary}

A search for heavy Majorana neutrinos in $\Pe^\pm \Pe^\pm \mathrm{jj}$ and $\Pe^\pm \mu^\pm \mathrm{jj}$ 
events has been performed using 19.7\fbinv of data collected during 2012 in $\Pp\Pp$ collisions at a centre-of-mass energy of 8\TeV. 
No  excess of events compared to the expected standard model background prediction is observed. Upper limits at 95\% CL are set on
$|\VeN|^2$ and $\abs{\VeN \VmNst}^2 / ( |\VeN|^2 + |\VmN|^2 )$  as a function of \mN in the range $\mN =40$-$500\GeV$, 
where $\VlN$ is the mixing element of the  heavy neutrino \N with the standard model neutrino $\nu_\ell$.

A significant increase in sensitivity on the limits for $\abs{ \VeN }^2$ has been achieved over the 
previous limits set by CMS with the 7\TeV data. These limits are the most restrictive direct limits for heavy Majorana neutrino masses above 200\GeV. 
The limits on $\abs{\VeN \VmNst}^2 / ( |\VeN|^2 + |\VmN|^2 )$ presented here are the first direct limits on this quantity for \mN above 40\GeV. 
For $\mN = 90\GeV$ the limits are $|\VeN|^{2} < 0.020$ and $\abs{\VeN \VmNst}^2 / ( |\VeN|^2 + |\VmN|^2 ) < 0.005$. 
At $\mN = 200\GeV$ the limits are $|\VeN|^{2} < 0.017$ and $\abs{\VeN \VmNst}^2 / ( |\VeN|^2 + |\VmN|^2 ) < 0.005$, 
and at $\mN = 500\GeV$ they are $|\VeN|^{2} < 0.71$ and $\abs{\VeN \VmNst}^2 / ( |\VeN|^2 + |\VmN|^2 ) < 0.29$. 

\clearpage

\begin{acknowledgments}

We congratulate our colleagues in the CERN accelerator departments for 
the excellent performance of the LHC and thank the technical and 
administrative staffs at CERN and at other CMS institutes for their 
contributions to the success of the CMS effort. In addition, we 
gratefully acknowledge the computing centres and personnel of the 
Worldwide LHC Computing Grid for delivering so effectively the computing 
infrastructure essential to our analyses. Finally, we acknowledge the 
enduring support for the construction and operation of the LHC and the 
CMS detector provided by the following funding agencies: BMWFW and FWF 
(Austria); FNRS and FWO (Belgium); CNPq, CAPES, FAPERJ, and FAPESP 
(Brazil); MES (Bulgaria); CERN; CAS, MoST, and NSFC (China); COLCIENCIAS 
(Colombia); MSES and CSF (Croatia); RPF (Cyprus); MoER, ERC IUT and ERDF 
(Estonia); Academy of Finland, MEC, and HIP (Finland); CEA and 
CNRS/IN2P3 (France); BMBF, DFG, and HGF (Germany); GSRT (Greece); OTKA 
and NIH (Hungary); DAE and DST (India); IPM (Iran); SFI (Ireland); INFN 
(Italy); MSIP and NRF (Republic of Korea); LAS (Lithuania); MOE and UM 
(Malaysia); CINVESTAV, CONACYT, SEP, and UASLP-FAI (Mexico); MBIE (New 
Zealand); PAEC (Pakistan); MSHE and NSC (Poland); FCT (Portugal); JINR 
(Dubna); MON, RosAtom, RAS and RFBR (Russia); MESTD (Serbia); SEIDI and 
CPAN (Spain); Swiss Funding Agencies (Switzerland); MST (Taipei); 
ThEPCenter, IPST, STAR and NSTDA (Thailand); TUBITAK and TAEK (Turkey); 
NASU and SFFR (Ukraine); STFC (United Kingdom); DOE and NSF (USA). 

Individuals have received support from the Marie-Curie programme and the 
European Research Council and EPLANET (European Union); the Leventis 
Foundation; the A. P. Sloan Foundation; the Alexander von Humboldt 
Foundation; the Belgian Federal Science Policy Office; the Fonds pour la 
Formation \`a la Recherche dans l'Industrie et dans l'Agriculture 
(FRIA-Belgium); the Agentschap voor Innovatie door Wetenschap en 
Technologie (IWT-Belgium); the Ministry of Education, Youth and Sports 
(MEYS) of the Czech Republic; the Council of Science and Industrial 
Research, India; the HOMING PLUS programme of the Foundation for Polish 
Science, cofinanced from European Union, Regional Development Fund; the 
OPUS programme of the National Science Center (Poland); the Compagnia di 
San Paolo (Torino); MIUR project 20108T4XTM (Italy); the Thalis and 
Aristeia programmes cofinanced by EU-ESF and the Greek NSRF; the 
National Priorities Research Program by Qatar National Research Fund; 
the Rachadapisek Sompot Fund for Postdoctoral Fellowship, Chulalongkorn 
University (Thailand); the Chulalongkorn Academic into Its 2nd Century 
Project Advancement Project (Thailand); and the Welch Foundation, 
contract C-1845. 

\end{acknowledgments}

\clearpage

\bibliography{auto_generated}

\cleardoublepage \appendix\section{The CMS Collaboration \label{app:collab}}\begin{sloppypar}\hyphenpenalty=5000\widowpenalty=500\clubpenalty=5000\input{EXO-14-014-authorlist.tex}\end{sloppypar}
\end{document}

%% file: EXO-14-014-authorlist.tex
\textbf{Yerevan Physics Institute,  Yerevan,  Armenia}\\*[0pt]
V.~Khachatryan, A.M.~Sirunyan, A.~Tumasyan
\vskip\cmsinstskip
\textbf{Institut f\"{u}r Hochenergiephysik der OeAW,  Wien,  Austria}\\*[0pt]
W.~Adam, E.~Asilar, T.~Bergauer, J.~Brandstetter, E.~Brondolin, M.~Dragicevic, J.~Er\"{o}, M.~Flechl, M.~Friedl, R.~Fr\"{u}hwirth\cmsAuthorMark{1}, V.M.~Ghete, C.~Hartl, N.~H\"{o}rmann, J.~Hrubec, M.~Jeitler\cmsAuthorMark{1}, A.~K\"{o}nig, M.~Krammer\cmsAuthorMark{1}, I.~Kr\"{a}tschmer, D.~Liko, T.~Matsushita, I.~Mikulec, D.~Rabady, N.~Rad, B.~Rahbaran, H.~Rohringer, J.~Schieck\cmsAuthorMark{1}, R.~Sch\"{o}fbeck, J.~Strauss, W.~Treberer-Treberspurg, W.~Waltenberger, C.-E.~Wulz\cmsAuthorMark{1}
% ...
\vskip\cmsinstskip
% ...
1:~~Also at Vienna University of Technology, Vienna, Austria\\
% ...
